# Solar System Observations with the James Webb Space Telescope

James Norwood
Dept. of Astronomy
Box 30001/MSC 4500
New Mexico State University
Las Cruces, NM 88003-0001
jnorwood@nmsu.edu

Heidi Hammel
Association of Universities for Research in Astronomy
1212 New York Avenue NW, Suite 450
Washington, DC 20005
hbhammel@aura-astronomy.org

Stefanie Milam
NASA Goddard Space Flight Center
Astrochemistry Laboratory
Code 691.0
8800 Greenbelt Rd.
Greenbelt, MD 20771
stefanie.n.milam@nasa.gov

John Stansberry
Space Telescope Science Institute
3700 San Martin Drive
Baltimore, MD 21218
jstans@stsci.edu

Jonathan Lunine
Carl Sagan Institute
402 Space Sciences
Cornell University
Ithaca, NY 14853
jlunine@astro.cornell.edu

Nancy Chanover
Dept. of Astronomy
Box 30001/MSC 4500
New Mexico State University
Las Cruces, NM 88003-0001
nchanove@nmsu.edu

Dean Hines
Space Telescope Science Institute
3700 San Martin Drive
Baltimore, MD 21218
hines@stsci.edu

George Sonneborn
NASA Goddard Space Flight Center
Code 665
Greenbelt, MD 20771
george.sonneborn-1@nasa.gov

Matthew Tiscareno
Center for Radiophysics and Space Research
Cornell University
Ithaca, NY 14853
matthewt@astro.cornell.edu

Michael Brown
Division of Geological and Planetary Sciences
California Institute of Technology
Pasadena, CA 91125
mbrown@caltech.edu

Pierre Ferruit
European Space Agency
European Space Research and Technology Centre
Keplerlaan 1, 2200AG
Noordwijk, The Netherlands
pferruit@rssd.esa.int

Abstract

The James Webb Space Telescope will enable a wealth of new scientific investigations in the near- and mid-infrared, with sensitivity and spatial/spectral resolution greatly surpassing its predecessors. In this paper, we focus upon Solar System science facilitated by JWST, discussing the most current information available concerning JWST instrument properties and observing techniques relevant to planetary science. We also present numerous example observing scenarios for a wide variety of Solar System targets to illustrate the potential of JWST science to the Solar System community. This paper updates and supersedes the Solar System white paper published by the JWST Project in 2010 (Lunine et al., 2010). It is based both on that paper and on a workshop held at the annual meeting of the Division for Planetary Sciences in Reno, NV in 2012.

**Introduction**

The James Webb Space Telescope (JWST) will succeed the Hubble Space Telescope as NASA's premier space-based telescope for planetary science and astrophysics. This 6.5-meter telescope, which is optimized for observations in the near- and mid-infrared portions of the electromagnetic spectrum, will be equipped with four state-of-the-art imaging, spectroscopic, and coronagraphic instruments. These instruments, along with the telescope's moving target capabilities, will enable the study of a vast array of Solar System objects with unprecedented detail, unhindered by the telluric features that afflict ground-based observers.

The intent of this white paper is to provide the latest information regarding JWST instrument sensitivities and capabilities for planetary science applications. We also present a number of hypothetical Solar System observations as a means of demonstrating potential planetary science observing scenarios, although the list of applications discussed here is far from comprehensive. The goal of disseminating this updated information is to stimulate discussion and participation among members of the planetary science community, many of whom are expected to become eventual users of JWST, and to encourage feedback on any desired capabilities that would enhance the usage of JWST for Solar System observations. JWST hardware is essentially complete at this time and is undergoing integration and cryogenic testing; the science instruments in particular are completely finished and no changes to them or their components are possible. However, the flight software capabilities for utilizing the observatory and instruments, and ground-system capabilities for planning observations and processing the science data are still being developed and will continue to evolve even after launch. These software systems can be adapted to achieve improved science return (within schedule and cost constraints), and community input on how they can be improved is timely and useful to the JWST project.

The number of possible types of Solar System observations has grown with further consideration by the JWST Science Working Group, by the increased participation of the community through workshops, and by the successful implementation of the moving target tracking. For this reason, we do not attempt an encyclopedic survey of all possible targets and observational types, but rather consider exemplary types of observations that demonstrate the substantial capability of the observatory in furthering Solar System research. It is our hope that readers will be sufficiently stimulated by the examples herein that they will develop their own proposals that might be submitted under general observing time.

**Motivation for Planetary Science Observations with JWST**

Numerous planetary science investigations will be enabled with JWST. The near- and mid-infrared spectral coverage and sensitivity afforded by JWST complements NASA's other Solar System exploration platforms such as Earth-based telescope facilities—both ground-based and in Earth orbit—and interplanetary spacecraft that continue to explore the Solar System through orbiter, flyby, and lander/rover missions. When JWST joins these important components of NASA's planetary exploration portfolio, it will offer improvements in sensitivity, spatial resolution, spectral resolution and coverage, and/or geographic area of exploration. JWST will contribute to the overarching objectives of planetary science, namely to understand planet formation, evolution, and the suitability of planets as habitats, through high-fidelity infrared imaging and spectroscopy of both large and small bodies in the Solar System. The suite of Solar System observations that will be enabled by JWST will advance our understanding of our own planetary system as well as more general astrophysical processes such as planet formation and evolution. A key role for JWST will be in characterizing the composition of objects that have previously been too small and/or too distant for such measurements. The ability to study the reflected and thermal spectrum of targets throughout the 0.7 – 28.5 micron region, un-impeded by the Earth's atmosphere, at unprecedented sensitivity and with a diffraction limited 6.5 meter primary mirror offers the possibility of ground-breaking discoveries in Solar System science.

Furthermore, if recent history is any indication, the number of known exoplanetary systems will continue to grow rapidly in the future, along with our ability to characterize exoplanets. Our own Solar System serves as a benchmark against which all other planetary systems are compared, and the insights gained into the physical processes governing the formation and evolution of our own system are by extension also relevant to the burgeoning field of exoplanet science. For example, the same dynamical processes manifest in the atmospheres of the giant planets of our own Solar System through cloud motions and signatures of trace molecular species are likely at work in extrasolar giant planet atmospheres. The study of hydrological, carbon dioxide, and methane cycles of our terrestrial planets and Titan will yield important insights into what we might expect to observe once the quest for Earth-like exoplanets transitions from discovery to characterization. The analysis of small bodies in our Solar System, which provide a window to the early stages of planetesimal formation from the protosolar nebula, will provide clues to the general formation processes occurring in stellar disks throughout our galaxy. In summary, there are numerous investigations of Solar System targets that can be enabled through JWST observations. These studies will not only advance our understanding of planetary formation, evolution, and habitats, but will also inform future exoplanetary investigations of sister worlds that have only recently been discovered.

**Special Needs for Solar System Observations**

Capabilities for Solar System Observations

Observations of various Solar System bodies can be conducted with JWST, and the science working group has put forth a significant effort in establishing the capabilities of the observatory for moving targets and bright objects. As the instruments continue being integrated with the observatory and the full system characterized, more accurate assessments of their capabilities are being made and will be updated through commissioning. Observation planning tools, which draw on existing capabilities developed to support Hubble Space Telescope (HST) observers, and associated capabilities are also being developed with the needs of Solar System observers in mind. Improvements include visualization of moving target tracks on the sky and automatic retrieval of orbital elements from NASA's HORIZONS ephemeris service operated by the Jet Propulsion Laboratory.

Moving Target Capability

JWST is designed to observe Solar System objects having apparent rates of motion up to 30 milliarcseconds/second. This capability includes the planets, satellites, asteroids, trans-Neptunian objects, and comets beyond Earth's orbit (see Table 1). The 30 mas/sec (3.6"/hr) speed limit will have small negative impacts for cometary science, and significantly larger impacts for NEO science. For known comets passing through JWST field of regard in the 2019 - 2020 time frame, over 95% can be tracked on any given day. For NEOs passing through the field of regard during that interval at least 80% can be tracked on any given day. For both comets and NEOs, the apparent motion preferentially exceeds the 30 mas/sec limit when the objects are nearest the observatory, and to the Sun. More detailed studies of these impacts will be undertaken in the near future. Moving (and fixed) targets can only be observed when they fall within JWST's field of regard (solar elongation of 85° to 135°, and a roll range of ±5° about the telescope's optical axis). The size of the field of regard is dictated by the thermal design of the observatory, specifically the 21 m × 14 m sunshield. By keeping the telescope and instruments in always the sunshield's shadow they are passively cooled to $T \sim 40$ K.

During the observation of a moving target, the science target is held fixed in the desired science aperture by controlling the guide star to follow the inverse of the target's trajectory. The ground system and on-board pointing control software use polynomial expansions of the target ephemeris. The JWST guider field of view (2.2′ × 2.2′) is located in the telescope focal plane several arcminutes from the science apertures. The predicted pointing stability for moving targets is <0.01″ (1σ radial), very similar to that for fixed targets.

Event-Driven Operations

On-board software scripts autonomously control the execution of the JWST science timeline (Balzano and Zak 2006). Previous space observatories operated using an observation schedule in which each on-board activity occurred at a specific time. JWST operations are event driven, meaning that activities are executed according to a *sequence*, but are not required to execute at specific times. This more flexible system should provide significant gains in efficiency. For example, the observatory will react to failed activities by immediately proceding to the next activity in the sequence rather than remaining idle until the scheduled time for the subsequent activity. The activity sequence is derived from proposals that support science, calibration, and necessary engineering activities. On-board, the activity sequence is composed of a series of "visits"; the relationship of visits to and the observing sequence to the proposals is as follows.

Proposals (whether for science, calibration, or engineering) are composed of one or more observations. An *observation* is a series of exposures with a single instrument to achieve a science objective. A *visit* is a series of exposures obtained with one science instrument and a single guide star (within the 2.2′ × 2.2′ field of the guider). An *observation* may be split into more than one *visit* (the *visit* is basic unit of scheduling) depending on its duration, distribution of guide stars, and other factors. The scripts respond to actual slew completion or on-board command execution, making operations more efficient. Scripts also respond to an interrupted or a failed *visit*, moving on to the next valid *visit*. *Visits* are scheduled with overlapping windows to provide execution flexibility and to avoid lost time. Each visit has an *earliest start time*, a *latest start time*, and a *latest end time* that define when each visit can be scheduled and executed. These timing parameters are determined from the program's scientific objectives. For visits without specific constraints on when they must execute, the interval between *earliest* and *latest start times* is expected to be a significant fraction of a day.

An observing plan covering about ten days will be uplinked weekly, but plan updates could be more frequent if necessary (for example, to accommodate a Target of Opportunity (ToO) observation).

The event-driven operations system does support time-critical observations, high-cadence time-series, monitoring, and Target of Opportunity (ToO) observations. The minimum response time for ToOs is 48 hours (observation approval to execution). Time-critical observations (i.e. those that must execute at a fixed time to achieve the scientific objective) can be specified with an uncertainty of 5 minutes; observers must accommodate this uncertainty in the design of their observations. For observations that must start within a timing window the system will guarantee that the observation begins within 5 minutes of the specified start time and no later than the end of the timing window (which must be at least 5 minutes after the start of the timing window). These capabilities are needed by a wide range of science programs, not just Solar System science.

Brightness Limits

The observed brightness of light reflected from a Solar System object is directly proportional to the area of the telescope's mirror and the angular size and albedo of the object, and inversely proportional to the square of the distance of the object from observer and the Sun. At thermal wavelengths the same is true except that the brightness depends on the object's temperature which scales as $(1-A)^{0.25}$, where $A$ is the bolometric albedo. The high sensitivity and large aperture of JWST are ideal for the smaller Solar System bodies such as KBOs and distant comets. In contrast, the outer Solar System planets are extremely bright, limiting the observatory's capabilities for conducting observations on these targets. Venus and Mercury cannot be observed due to the solar elongation limits on JWST's pointing (limited from 85° to 135°). Below we present a few cases where outer-planet observations are feasible based on preliminary knowledge of observatory and instrument performance; the optimal observing strategies are summarized in Table 3. Note that this is not inclusive of all Solar System studies that can be conducted, and special cases (modes, target acquisition, etc.) may exist for some extreme targets. Where there is disagreement, results presented in this study supersede those in the JWST technical report by Meixner et al. 2008 (JWST-STScI-001375).

The four JWST science instruments provide imaging, coronagraphy, and spectroscopy over the 0.6–28.5 μm wavelength range. (See Gardner et al. (2006) for instrument design details.) Imaging fields of view are ~2′ × 2′, with pixel sizes of 0.032″ (0.6-2.5 μm), 0.065″ (2.5-5 μm), and 0.11″ (5-28.5 μm). The spectroscopic capabilities include 1-5 μm multi-object spectroscopy over 3′ × 3′ with 250,000 individually addressable shutters (each 0.2″ × 0.46″), integral field spectroscopy covering 1-28.5 μm with a field of view of 3″ × 3″ or larger, and several long slits. The spectrographs provide spectral resolving powers of ~100 to ~2700 over the 1-28.5 μm range.

NIRCam

The minimum time to read out a NIRCam detector in full-frame mode is 10.74 seconds. Mars, Jupiter, and Saturn will saturate in all NIRCam filters in full-frame mode, and Uranus and Neptune will saturate at the shorter wavelengths. Subarray performance in this application, e.g. against a bright extended source, has not been confirmed. Assuming that this mode works satisfactorily, subarrays will provide significantly shorter integration times, and make it possible to observe these objects without saturating the detectors. Figure 1 illustrates the configuration of subarrays in the NIRCam short wavelength (SW) channel (0.6-2.5 μm) for imaging of extended sources. Gaps between the four detectors in the SW focal plane would be filled by dithering. The longwave (LW) channel has a single detector with a plate-scale twice that in the SW channel, and is automatically configured such that the LW FOV is matched to the footprint of all 4 SW FOVs. With 640×640 pixel subarrays the FOV is 41.6″ across, large enough to encompass the disk of Jupiter, as shown in Figure 1.

The 1-5 μm spectra of the outer planets are compared with the NIRCam saturation limits (for $640^2$ subarrays) in Figure 2. Uranus and Neptune can be observed in all NIRCam filters without saturating. Jupiter and Saturn exceed the saturation limits shortward of about 1.5 μm, although it may be possible to observe them using the medium bandwidth filters F140M and longer; in the longwave channel they can be observed in many of the filters without saturating. Because its apparent size is about half that of Jupiter, Saturn could be observed using the $400^2$ subarray configuration (with saturation limits 2.5 times higher); that would allow for un-saturated images in most NIRCam filters. Mars is so bright that it can only be observed in the narrow-band filters except at 4.3 μm. The $400^2$ subarray doesn't increase the saturation limits enough to change that, but Mars is small enough to nearly fit within a $160^2$ subarray, where the saturation limit is 16 times greater than shown in Figure 2. The $160^2$ subarray is available in point-source photometry mode, and by executing dithers with a throw of a few arcseconds the entire disk could quickly be mapped.

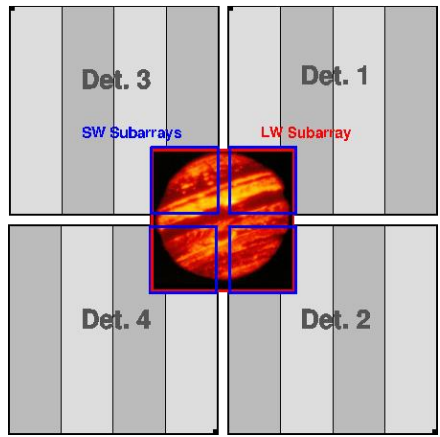

Figure 1. Proposed configuration for observing bright targets with NIRCam subarrays.

Jupiter (and most of the bright objects in the Solar System) can be observed if the object is centered between the 4 detectors allowing for four subarrays with a 40″ FOV (see Figure 1) across both the short wavelength and long wavelength channels (Figure 2).

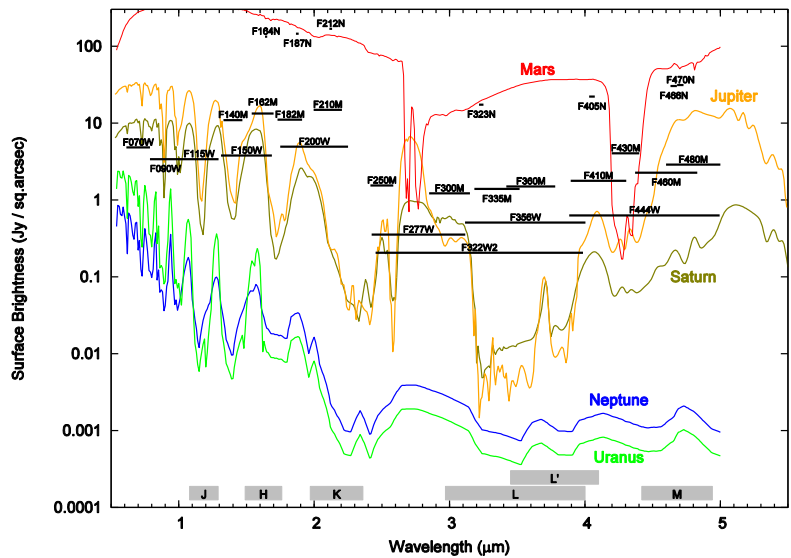

Figure 2. Saturation limits (black) for the NIRCam filters employing the subarray configuration in Figure 1, compared to planetary spectra. Spectra composited from McCord and Adams (1969), McCord and Westphal (1971), Lellouch et al. (2000), Clark and McCord (1979), Karkoschka (1994), Encrenaz (1997), Fink and Larson (1979), and Burgdorf (2008).

NIRSpec

The full-well capacity for the NIRSpec detectors is 77,000 electrons. Brightness limits for NIRSpec observations of the bright Solar System bodies were derived using our current knowledge of the instrument. Two different modes were considered: the IFU mode where a 3″ × 3″ field of view is observed with a sampling of 0.1″, and the 0.2″ × 3.3″ fixed-slit mode. These were then compared to each planet's surface brightness (Table 2) for various configurations and resolving powers, including the "Window" and "Stripe" sub-array modes. The results are summarized in Figures 3-4 and Table 3.

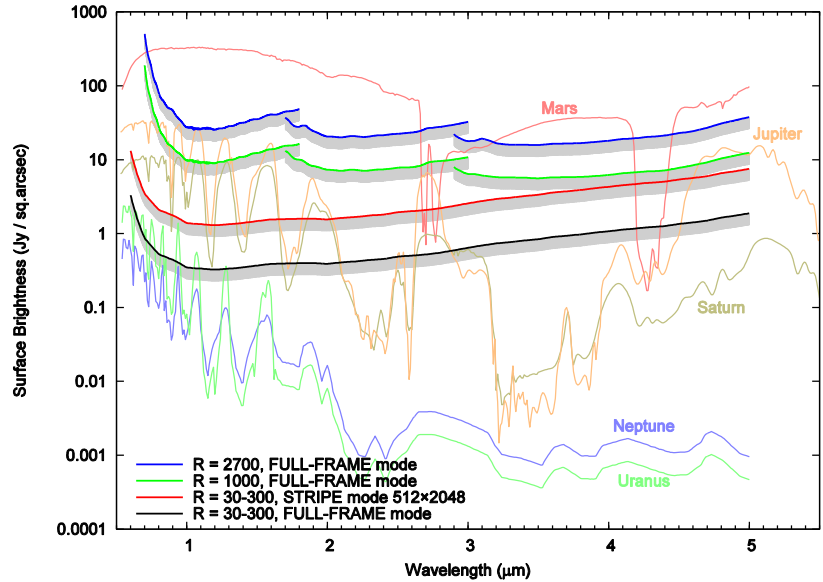

Figure 3. Maximum surface brightness limits for NIRSpec in IFU mode, assuming two frames per exposure, with a full well of 77,000 electrons. The gray zone below each curve highlights the range corresponding to a (not unrealistic) 30% uncertainty level on the limits. The planet spectra are the same as in Figure 2.

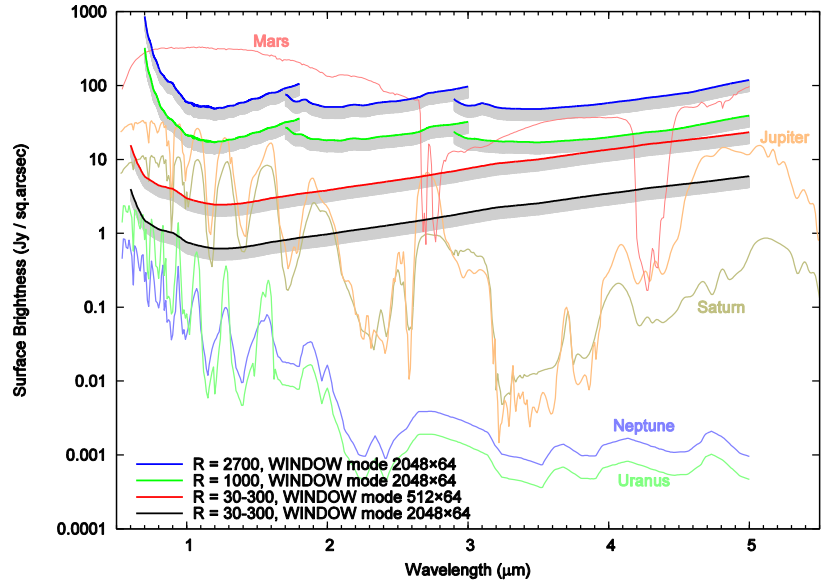

Figure 4. Maximum surface brightness limits for NIRSpec in slit mode, assuming two frames per exposure, with a full well of 77,000 electrons. The gray zone below each curve highlights the range corresponding to a (not unrealistic) 30% uncertainty level on the limits. The planet spectra are the same as in Figure 2.

MIRI

The MIRI instrument has nine broad-band imaging filters (with four defined subarrays, plus full field, to date), a medium-resolution spectrometer (MRS), and a low-resolution spectrometer (LRS). For bright extended objects, the SUB64 subarray (frame time of 0.085 s)[1] is the most promising approach for imaging. However, the performance of subarrays on bright sources has not yet been verified. If the capability to use subarrays on bright sources is confirmed in future tests, and assuming a full well depth[2] of 225,000 $e^-$ and a quantum efficiency midway between the specification and 100% (e.g., 70% at 5.6 μm, 80% around 12 μm, 75% around 20 μm), an estimate of the integration time to saturation can be made for each of the surface brightnesses presented in Table 2. In this SUB64 imaging mode, all the bright planets except Mars can be observed with at least one of the broad-band MIRI imaging filters (see Chen et al. 2010). The MIRI spectrometers offer more options for observations since they disperse the light of bright sources. Again, all the bright planets are accessible in at least half the available mid-infrared spectroscopic bandwidth except Mars.

NIRISS

---

[1] Two frames minimum are required to fit the slope and derive a count rate, so the minimum on-source integration time is 0.170 seconds.

[2] Note that the surface brightnesses presented in Table 2 are disk-averaged. In practice, planets may have varying features across the disk, so conservatively using an upper limit to the integration time of ~2/3 full-well may be more appropriate.

The NIRISS instrument has a larger pixel resolution than NIRCam, higher sensitivity to extended emission < 2.5 μm than NIRCam (but at lower spatial resolution). NIRISS can image with a full 2048×2048 pixel array, SUB256, SUB128 and SUB64 with readout times of 10.74 s, 0.70 s, 0.18 s, and 0.05 s, respectively, including seven wide filters and five medium filters. Currently, imaging is only offered from the full array since NIRCam will likely provide better performance in the subarray modes. Spectroscopy with NIRISS will be accessible with either the GR150 grism or the GR700 grism. The GR150 offers Wide-Field Slitless Spectroscopy with a resolving power of ~150 and spectral coverage of 0.8-2.2 μm. Sensitivity limits for this mode are presented in the specific examples to follow and allow for spectral observations of most objects in the solar system at a lower spectral resolution. Also of interest is unique mode of Aperture Masking Interferometery with NRM. This is a seven-hole aperature mask with 21 distinct baselines. While this mode was optimized for exoplanet detections around bright stars, case studies are being considered for solar system targets, such as binary asteroids.

**Example Observing Programs**

Mars

Studies of Mars with JWST offer the promise of unique and important new contributions to Mars science and to NASA's future mission goals. Specifically, global-scale near-IR observations can: (1) determine the variability of major and secondary atmospheric species like $CO_2$, CO, and $H_2O$, providing data for photochemical and dynamical modeling of the present Martian climate; (2) constrain the near-IR radiative and absorptive properties of airborne dust, another key component of the present Martian climate system; (3) help to quantify the surface volatile budget and resource potential by detecting and mapping the distribution of $H_2O$-bearing or OH-bearing surface minerals like clays and hydrates; and (4) assess the magnitude and scale of diurnal, seasonal, and interannual volatile transport through direct near-IR detection and discrimination of surface and atmospheric $H_2O$ and $CO_2$ ices/clouds, especially in the polar regions.

The orbital eccentricity of Mars ($e = 0.0933$) means that the insolation between perihelion and aphelion varies by ~40%, so any attempt to use seasonal symmetries is not a viable strategy for understanding seasonal transitions (e.g., spring equinox vs. fall equinox does not represent the same transition). Additionally, since observing epochs are tied to our synodic period with Mars rather than the planet's orbital period, most seasonal transitions will not cleanly coincide with observing windows. As a result, a complete picture of Mars' seasonal cycles will require observations obtained at multiple epochs and over extended, contiguous periods. Current and planned probes are not capable of sampling multiple local times (i.e., local diurnal variation) on sub-seasonal time-scales for most of the Martian surface area; they are not Mars-synchronous, and are positioned in high-inclination orbits. Furthermore, there are no planned thermal-IR instruments for Mars.

While HST affords a look at the entire disk, the current instrumentation is limited to wavelengths shorter than 1.6 μm. In addition, the orbital constraints on HST forbid prolonged, continuous monitoring of the atmospheric changes over a Martian day. JWST solves these problems with its suite of instruments, and its L2 orbit, which enables the observatory to monitor the Martian atmosphere on timescales from a few minutes to weeks, with any cadence that is required by the observer.

Mars is a very bright object for JWST: the planet will saturate in MIRI observations, and most spectral regions shortward of 2.65 μm. Near-infrared observations of Mars are possible with NIRSpec, mainly in the strong $CO_2$ absorption features near 2.7, 2.8, and 4.3 μm; longward of 3 μm, where extinction from dust and clouds diminishes the brightness of Mars; and near 0.7 μm, where the spectrometer's saturation limit is highest. Example investigations include synoptic monitoring of gases, aerosols, and dust in the Martian atmosphere over the entire disk. NIRSpec will be used to measure the strengths of ice features, which constrain particle size (aerosol components).  Due to the high surface brightness of Mars, observations of this planet with NIRSpec will only be possible beyond 2.5 μm and using the 0.2″ × 3.3″ slit at high spectral resolution (see Figure 4).  The primary atmospheric components measurable with NIRSpec are CO and $H_2O$.  NIRSpec would only be able to detect methane in concentrations about a factor of 5-10 larger than previous estimates and newer upper limits (e.g., Webster et al. 2013).

A comprehensive observing program could be conducted over two consecutive Martian days (24.6 Earth hours), four times during the Martian year.  The four yearly epochs will coincide with important seasonal transitions where the atmosphere is undergoing significant heat load changes.  It is necessary to monitor Mars for two full Martian days in order to get full coverage of the Martian surface.  This observation program also assumes Mars to be near aphelion to reduce the planet's brightness.

Jupiter and Saturn

The JWST era will be an important time period for extending, enhancing, and complementing the infrared observations of Jupiter and Saturn made in the preceding decades by Galileo, Cassini, Juno, New Horizons, Spitzer, and HST.  Jupiter is the largest, most massive, and closest of the gas giant planets, and thus has been the best studied with both ground-based and space-based (orbiter and flyby) assets.  Cassini is the longest orbiting mission in the outer Solar System, and has been observing many aspects of the Saturn system for nearly the past decade.  Despite these observational achievements, there are numerous outstanding questions about the giant planet atmospheres that can be addressed with JWST.

The relationships between heat transport, atmospheric dynamics, and chemical processes are fundamental to our understanding of the giant planet atmospheres.  Furthermore, Jupiter represents the archetype of a broad class of giant planets, many of which are now known to exist in orbit around other stars, and provides a nearby "laboratory" where we can test our understanding of giant planet formation and evolution.  Two example outstanding questions in the field of giant planet atmospheres, which are relevant to both our own gas giants as well as a myriad of exoplanets, include

- How does atmospheric circulation detected at the cloud tops relate to deeper atmospheric motions and heat transport from the interior? and
- What is the link between atmospheric dynamics, chemistry, and cloud microphysics?

Many components of these questions can be addressed with JWST observations of Jupiter and Saturn. The infrared spectra of Jupiter and Saturn are shaped by a combination of reflected sunlight and thermal radiation escaping from the deeper atmosphere, modulated by the spectral properties of the gases and aerosols present in the overlying cloud decks.  For example, the relationship between atmospheric chemistry and dynamics can be explored through the observation of disequilibrium species such as phosphine ($PH_3$) and arsine ($AsH_3$), and the spatial variations thereof.  These species are tracers of tropospheric mixing since their observed abundances exceed those predicted from thermochemical equilibrium calculations.  The chemical composition of giant planet atmospheres as well as cloud and aerosol properties can be determined using IR imaging and spectroscopy from JWST, as detailed below.

Near-Infrared Studies

The near-IR spectra of Jupiter and Saturn are dominated by methane absorption bands.  When imaged in these bands, the planets appear dark, thus any reflective aerosols must be located high enough in the atmospheres to be above most of the methane-absorbing gas.  Imaging in these bands therefore provides an opportunity to sound the upper troposphere or lower stratosphere and gain information about the variation of certain atmospheric properties as a function of altitude – e.g., by tracking cloud features to determine wind speeds.

Jupiter and Saturn are bright enough to saturate in some NIRCam filters under normal circumstances, though the use of sub-array imaging techniques can make all near-infrared filters useable. An efficient configuration has the planet centered between the four detectors, as shown in Figure 1; this can be accomplished with the $640^2$ (40″ × 40″) subarray for Jupiter, and the smaller $400^2$ (20″ × 20″) subarray for Saturn. For Jupiter, employing the $640^2$ subarray will make all narrow filters and most medium filters useable. Other filters will require the $160^2$ or even $64^2$ subarrays; since those will involve fields of view smaller than the Jovian disk, observers will need to focus on specific regions of interest, or employ mosaicking techniques to construct a full-disk image. Due to its generally lower brightness and smaller size, Saturn will be less difficult: almost all NIRCam filters will be useable with the Saturn-sized $400^2$ subarray, and the remaining wide filters (F080W, F115W, F77W, F322W2, and possibly F070W and F150W) will not saturate with the $160^2$ subarray.

Near-IR images of Jupiter and Saturn are shown in Figure 5. This spectral region covers a wide range of physical processes, including gaseous and aerosol absorption and scattering, thermal radiation, and cloud opacity. JWST will enable extensive imaging of the gas giant planets through filters that are sensitive to species such as ammonia and methane, providing insight into energy transport and the chemistry in these atmospheres.

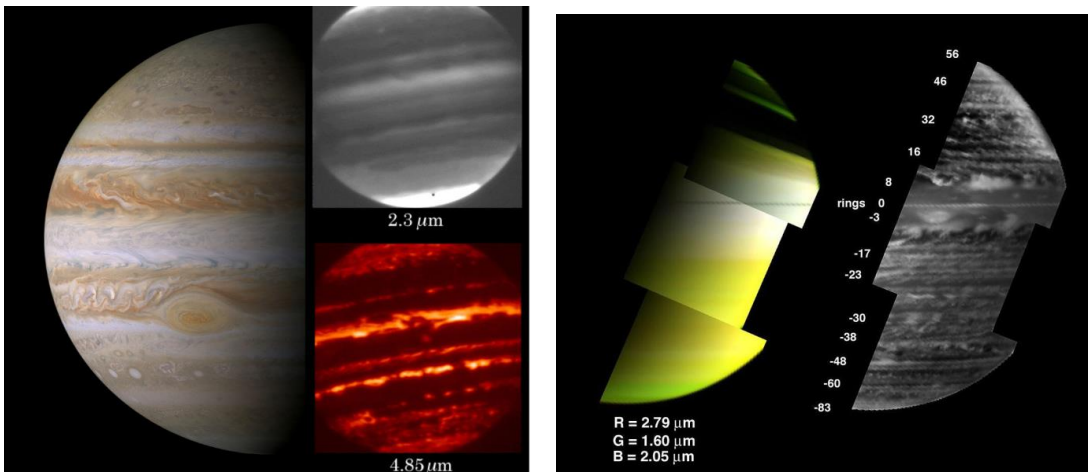

Figure 5. Left Panel: Jupiter's visible cloud deck is shown in the left image, taken by the ISS imager on the Cassini spacecraft in December 2000. The upper right image shows a ground-based image of Jupiter taken from NASA's Infrared Telescope Facility (IRTF) through a 2.3 μm filter, which is centered on a strong methane absorption band. The white clouds in this image are higher than the darker regions. The bottom right image shows an IRTF image of Jupiter at 4.85 μm, which is sensitive to thermal radiation from the lower cloud layers. In this image dark regions correspond to thicker clouds, whereas bright areas are relatively cloud-free. Right Panel: Saturn images taken by the Visual and Infrared Mapping Spectrometer (VIMS) on board the Cassini spacecraft. The left image shows a color composite sensitive to aerosols in Saturn's upper atmosphere, and the right image shows Saturn at 4.85 μm. Both images are from Sanchez-Lavega (2011), *An Introduction to Planetary Atmospheres*, Chapter 5.

Both Jupiter and Saturn can be observed with NIRSpec in fixed-slit mode (with a slit size of 0.2″ × 3.3″) or IFU mode (with a field of view of 3″ × 3″), and resolving powers of ~1000 and 2700. The 1-5 μm spectral region in the gas giants is quite rich, containing absorptions due to gaseous ammonia and methane, pressure-induced molecular hydrogen absorption, and disequilibrium species such as phosphine, arsine, and carbon monoxide (Figure 6). JWST will enable targeted spectroscopy of localized regions on Jupiter and Saturn, covering for example Jupiter's Great

Red Spot or a storm cloud on Saturn, which will provide unique data concerning the linkage between local chemistry in the gas giant atmospheres and larger scale atmospheric dynamics.

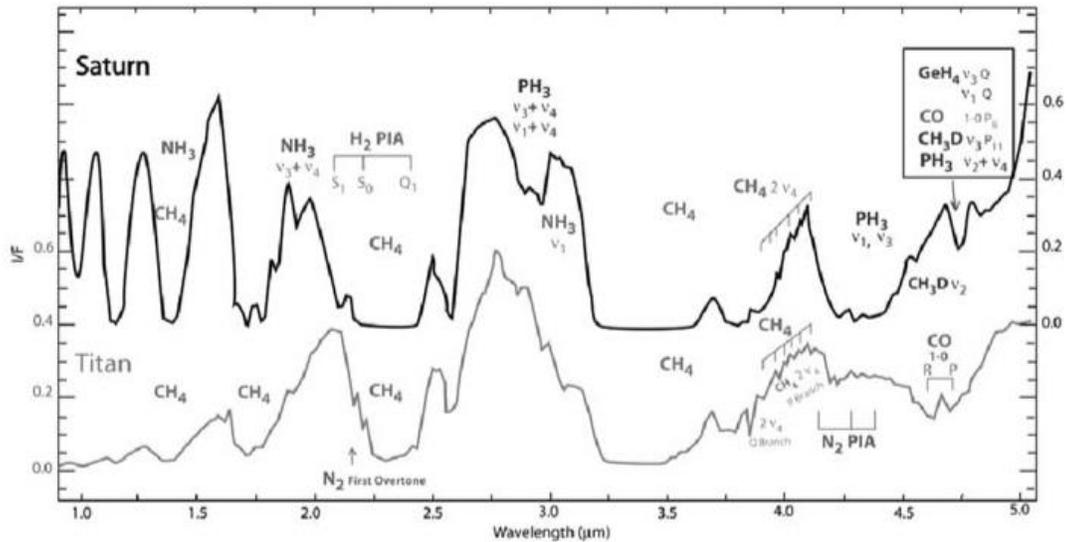

Figure 6. VIMS spectrum of Saturn (darker line) over a spectral region comparable to that of NIRSpec. The spectral resolution of the VIMS data is ~200; NIRSpec will be able to match or exceed this, providing a more detailed understanding of the chemistry and dynamics in the giant planet atmospheres. Figure from Baines et al. (2005).

Mid-Infrared Studies

As mentioned above, for bright extended sources such as the giant planets, imaging with MIRI is best accomplished in the SUB64 subarray mode, which yields a field-of-view size of 7″. Thus, in order to image the entire disk of either Jupiter or Saturn, multiple pointings will be required to create a full-disk mosaic. Figure 7 illustrates the surface brightness of Jupiter and Saturn along with the saturation limits of the MIRI filters. The Jupiter curves include both a hot spot, where the ammonia cloud deck is relatively thin, allowing one to see to regions in the planet dominated by Jupiter's thermal emission; and a zone, where the ammonia clouds are thicker and block the Jovian blackbody radiation. Jupiter's hot spots will likely be too bright for all the MIRI filters with the possible exception of F560W; the zones on Jupiter will be visible in the F560W filter as well as perhaps the F770W filter. Saturn will be observable in all MIRI filters shortward of 17 μm (i.e. F560W, F770W, F1000W, F1130W, F1280W, and F1500W).

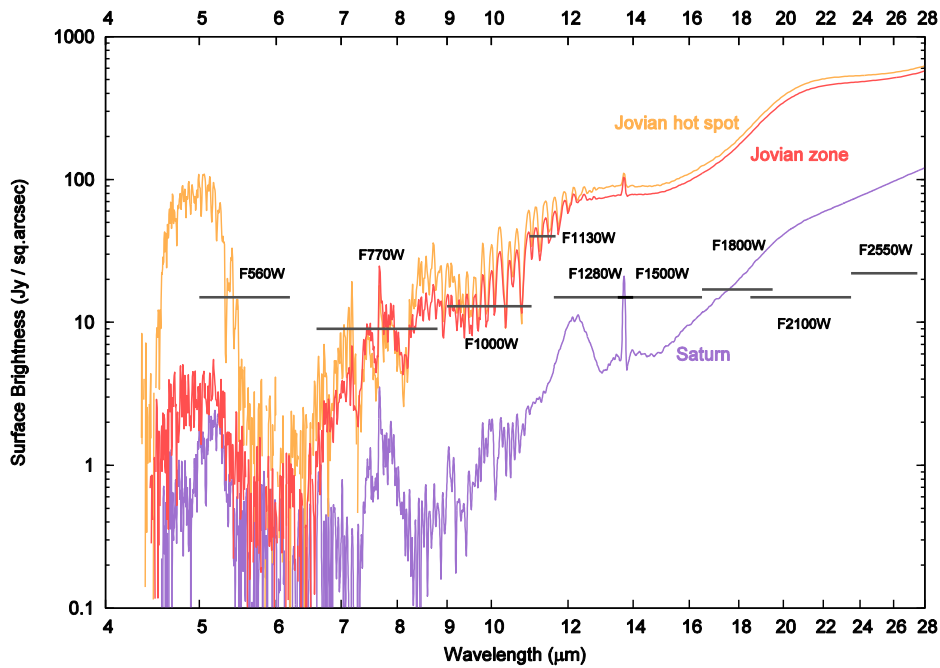

Figure 7. Spectra of a Jovian hot spot, a Jovian zone, and Saturn (G. Bjoraker, personal communication), compared to saturation limits of MIRI filters assuming SUB64 sub-array viewing with minimum integration time.

Spectra of the giant planets can be acquired with MIRI in several integral field unit (IFU) medium resolution spectroscopy modes shortward of ~10 μm (Jupiter) or ~17 μm (Saturn). As shown in Figure 7, the IR spectra of the gas giants are shaped by gaseous absorption and emission features such as that due to acetylene ($C_2H_2$) near 14 μm. The MIRI IFU's are significantly smaller than the angular sizes of Jupiter and Saturn (~3″ vs. ~20-40″), so they can be used for targeted observations of latitudes of particular interest, such as polar regions with stratospheric haze, convective cloud regions, and vortices such as the Great Red Spot.

Uranus and Neptune

The ice giants Uranus and Neptune are much more distant than Jupiter and Saturn, and as a result they have historically been greater challenges to observe them in detail. It was not until the debut of space telescopes and adaptive optics that Earth-based observations were finally able to resolve individual features on these two planets, such as latitudinal banding and isolated cloud features. In addition to serving as one of the few facilities with the spatial resolution for such observations, JWST will provide substantial improvements over existing capabilities. It has a larger collecting area than Hubble and Spitzer, allowing a greater signal-to-noise ratio in its observations; and it will expand spectral coverage into the mid-infrared, a region unavailable to Hubble and Keck.

With the greater sensitivity and spatial resolution afforded by JWST in the near- and mid-infrared, investigations can address some key questions regarding the ice giants, including the following:

- What are the global circulation patterns on these planets? If Uranus and Neptune exhibit different circulation morphologies, are the differences related to Uranus' unique seasonal insolation patterns, or to internal differences?

- How do the chemical and thermal structures of these planets' atmospheres vary with latitude? At what depth do they become horizontally uniform?

- In what ways do these planets' upper tropospheres interact with their stratospheres?

- To what extent does infalling material from rings, satellites, and/or comets affect the composition and chemical processes in these planets' stratospheres?

While most ground-based studies of Uranus and Neptune have been limited to full-disk observations, the high spatial resolution of JWST enables observation of how the features described above vary across the face of each planet. Knowing how the cloud layers, hydrocarbon distribution, and other properties change with latitude and longitude will greatly assist in the development of dynamical models that characterize these planets' atmospheric circulation. Repeated observations will provide temporal information as well. In addition to greatly improving our state of knowledge of Uranus and Neptune, such investigations will also be useful in comparative planetology, illuminating the properties that make the ice giants a unique class of planet, distinguishing them from the gas giants Jupiter and Saturn.

Near-Infrared Studies

Most features in the near-IR spectra of Uranus and Neptune arise from variability in the absorption spectrum of methane. The absorption strength of methane varies by several orders of magnitude, allowing observations to probe a wide variety of altitudes. In spectral regions where multiple $CH_4$ lines overlap for strong absorption, we see only light scattered at high altitude, primarily from stratospheric haze. On the other hand, in the windows with minimal $CH_4$ absorption, we see down to the upper cloud decks, understood to be placed between 1 and 10 bars. The precise altitudes of these cloud layers may be determined from spectral regions where $CH_4$ absorption is of intermediate strength, such as at the boundaries of the $CH_4$ windows. Furthermore, examination of different $CH_4$ windows will reveal properties of the clouds themselves: for example, as the wavelength increases beyond the size of a typical cloud particle, the cloud's ability to effectively scatter sunlight decreases.

For imaging the outer planets and other bright compact objects, sub-array imaging has been developed for NIRCam and MIRI in order to lessen the minimum integration time. However, while Uranus and Neptune would seem to be ideal subjects for sub-array imaging due to their small angular sizes, they are also dim enough for the technique to be unnecessary under most circumstances. Neptune will be observable in all NIRCam filters without subarrays; for Uranus subarrays may be needed for the F070W and F090W filters. NIRCam's pixel scales are 0.032″ between 0.6 and 2.3 µm, and 0.065″ between 2.4 and 5.0 µm: Uranus' diameter in these modes would be 112 and 55 pixels, while Neptune would be 72 and 35 pixels across.

Useful NIRCam filters include F182M, which spans the boundaries of windows in the absorption spectrum of methane; F335M, which covers an $H_3^+$ feature at 3.4 µm; and F480M, which catches a CO feature at 4.7 µm. Numerous features also span boundaries of methane absorption windows, allowing identification of vertical structure. At longer wavelengths, the planets' cloud decks become more optically thin, enabling views deeper within their atmospheres.

NIRSpec is capable of observing Uranus and Neptune from 0.6 to 5.0 microns at various spectral resolutions ($R$ ~100, 1000 and 2700). The most interesting option is to use the IFU to observe these planets. In this mode, a total of 900 spectra can be obtained simultaneously, covering a 3″ × 3″ field of view, with spatial sampling of 0.1″. Uranus' angular diameter of 3.6″ will be larger than the IFU's field of view, but full coverage may be obtained with mosaicking. An observer may also achieve better spatial resolution by dithering observations.

Mid-Infrared Studies

In the mid-IR, the reflectance spectra for these planets are replaced by thermal emission. In this spectral region the dominant opacity source is collisional absorption between $H_2$ and He. Also present are numerous emission features due to $CH_4$, $CH_3D$, and higher-order hydrocarbons created through photochemistry in the stratosphere. Mid-IR observations with JWST can greatly improve our understanding of these planets' thermal structure, D/H ratio, and photochemical processes.

As discussed in an earlier section, the smallest sub-array available to MIRI will be the $64 \times 64$ sub-array, which has a field of view 7″ across, twice Uranus' angular diameter. Shown in Figure 8 are the surface brightnesses of Uranus and Neptune compared to the saturation limits using the SUB64 sub-array. Both planets will be observable in all MIRI filters with sub-array imaging.

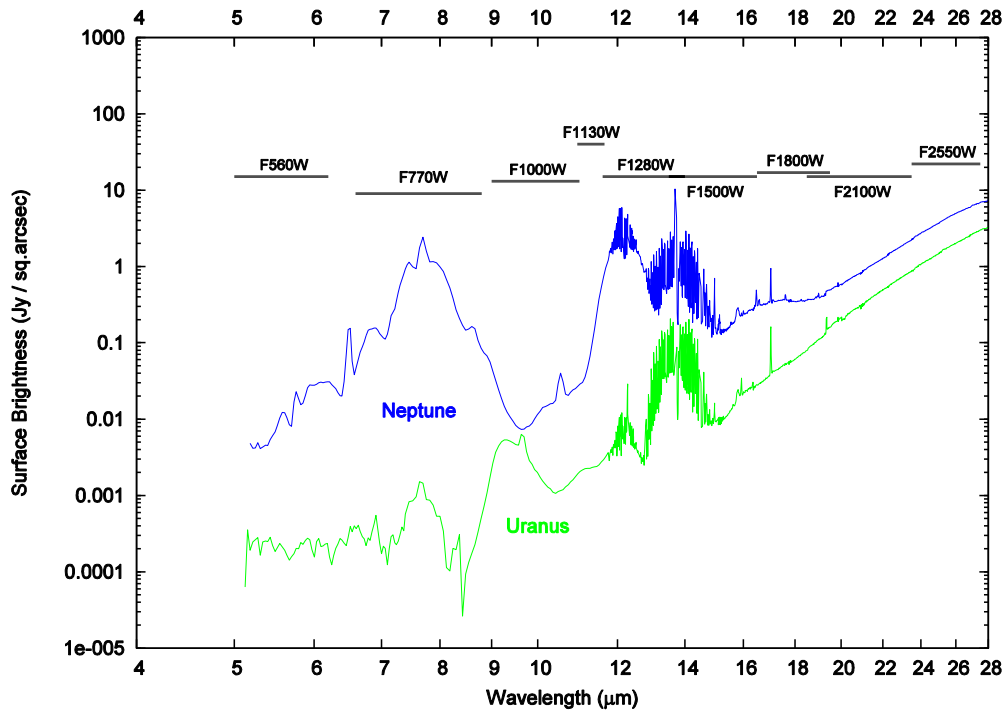

Figure 8. Spitzer/IRS spectra of Neptune and Uranus (Orton et al. 2014) compared to saturation limits of MIRI filters, assuming SUB64 sub-array imaging with minimum integration time (a factor of ~33 shorter than without sub-arrays). Neptune would be observable without sub-arrays in F560W, F1000W, F1130W, and possibly F1800W. Uranus would only require sub-arrays for F2100W and F2550W.

MIRI's Medium-Resolution Spectrograph (MRS) has four integral field units that are able to provide spatially resolved spectra across Uranus and Neptune, as shown in Figure 9. For example, the shortest-wavelength unit has a field of view 3.7″ on a side, which is divided into 30 strips for spectra. Including the pixel scale of the spectra, the IFU effectively divides the field into $0.18″ \times 0.19″$ regions: about twenty across the Uranian disk, and 12 across the Neptunian disk. The longer-wavelength IFU's have lower spatial resolution, with the poorest still offering $13 \times 5.5$ resolution elements across Uranus, and $8.5 \times 3.5$ across Neptune. This spatial resolution may be improved by a factor of 2 with four-position dithering.

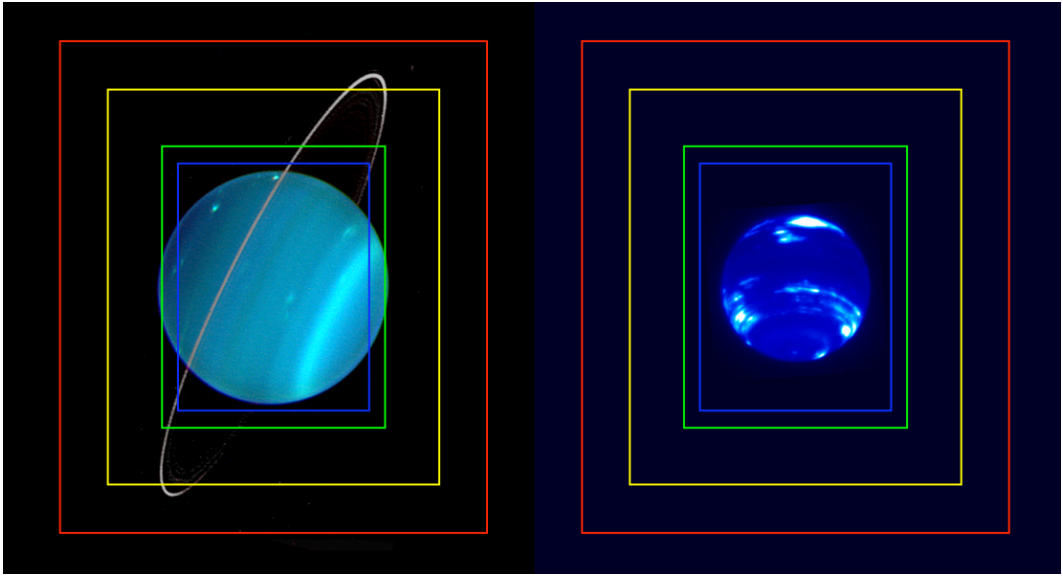

Figure 9: Sample viewing geometry for observing Uranus and Neptune with the Medium-Resolution Spectrograph IFUs. Shown are the fields of view for the four different IFUs compared to the typical sizes and orientations of Uranus and Neptune during JWST operations. Uranus image credit: L. Sromovsky and W. M. Keck Observatory; Neptune image credit: I. de Pater, H. B. Hammel, and W. M. Keck Observatory.

Europa

Jupiter's satellite Europa contains what is likely the largest reservoir of accessible liquid water anywhere in the Solar System other than the Earth itself (Pappalardo et al. 1998). With energy input from a warm radioactively decaying chondritic-composition core and an intense external radiation environment, chemical reactions in the internal ocean might be capable of building some of the pre-biotic materials thought to be necessary for life (Hand et al. 2007). But even after almost a decade of intense scrutiny from the Galileo spacecraft, debate still persists about the nature of the surface chemistry and the relative roles of exogenous radiation processing versus endogenous oceanic emplacement. From the current data, Europa can be viewed as a purely passive ice shell onto which ion and electron bombardment creates a limited chemical cycle confined to a thin surface layer, or it can be seen as a geologically active body with a chemically rich ocean and a sulfur-rich surface, both of which feed each other and record a complex chemical cycle.

The key to determining the nature of Europa is the composition of the surface and whether or not chemicals expected from an oceanic source can be seen on the satellite. The surface of the trailing side of Europa, in particular, is dominated by some sort of hydrated compound, rather than more pure water ice seen on the leading side, but the precise composition of this dominant chemical component on the trailing side of Europa is still unresolved. Much of the reason for the chemical uncertainty is the spectral similarity of many of the proposed constituents, particularly in the well-observed 1-2.5 µm range where spectra of hydrated materials are dominated by the strong absorptions due to their water components and the lack of diagnostic spectral features out to 5 µm observed by the Galileo spacecraft.

It is in the mid-infrared that JWST will best contribute to Europa science. Europa and the other icy Galilean satellites are poorly characterized beyond 5 µm, and in particular between 5 and 15 µm where Voyager obtained few results. This spectral region contains anion-specific combination bands of proposed constituents of the icy satellite surfaces, key spectral features that are diagnostic of the radiolytic cycle, and the potential for unexpected discoveries.

On Europa, many of the most important potential spectral features lie in the 5-10 µm region. A series of MIRI spectra covering this wavelength range and sampling ~4 regions around the satellite would allow several key scientific investigations, including the following:

- Searching for the strong spectral signature of hydrated minerals (Figure 10). If the surface of the trailing side of Europa is rich in ocean-derived hydrated minerals, their spectral signature could dominate the 5-7 µm region. Radiolytic-produced sulfuric acid, in contrast, is essentially featureless in this spectral range. Spectral differences between the leading water ice rich hemisphere and the trailing hydrate rich hemisphere could be particularly illuminating. The clear detection of hydrated minerals would demonstrate a strong connection between the interior ocean and surface, and point to a rich chemistry in the interior and on the surface of Europa.

- Understanding the importance of a radiolytic carbon cycle. To date, it is known that water and sulfur participate in the radiolytic cycle, but no evidence for carbon-based radiolysis has been found in spite of the presence of $CO_2$ in the surface ices of Europa (Carlson et al. 2009). Laboratory experiments (Figure 11) have shown strong carbonic acid features at 5.83 µm, 6.63 µm, and 7.65 µm when simulated Europa ices are bombarded with electrons (Gerakines et al., 2000; Hand et al., 2007). If Europa's surface is rich in carbonic acid and has an active ice shell that exchanges material with the ocean, then Europa's ocean might not only be oxygen-rich (Chyba, 2000), but it could be alkaline and contain a biologically useful source of carbon (Hand et. al., 2007).

- Searching for carbonyls (C=O) and amides (H-N-C). These may be detected by observing the broad 6-7 µm water ice feature, whose shape is modified by these species (Figure 11). Nitrogen in particular, has never been detected at Europa and could be an important unexplored component of the satellite's current ocean or surface chemistry.

- Searching for possible organic features (C-C, C-H, C=C) in the 6-8 µm region. Organics have been detected on Ganymede and Callisto in the 3.44-µm C-H region (McCord et al., 1998), but to date no organics have been observed on Europa. Laboratory experiments show that under Europan conditions, the radiolytic processing of hydrocarbons produces distinct absorptions in the 6-8 µm range (Hand and Carlson, 2011). In particular, methane is produced and trapped in the ice. Recent results in the 3.44-µm region with Keck spectroscopy (Brown and Hand, in prep) suggest that detection of organics is unlikely; nonetheless, the importance of organic chemistry warrants a search for these features on Europa.

- Detection of unexpected spectral features. While laboratory experiments have attempted to simulate all of the relevant chemistry on Europa, much remains unknown. The discovery of any new features could be indicative of a major chemical pathway previously unknown on Europa and the other satellites.

For a bright object as well-studied as Europa, few opportunities exist to observe completely unseen spectral regions. It is thus no surprise that a first look at this spectral region could potentially yield so rich a bounty. Accomplishment of any one of the scientific objectives here has the chance of achieving a major breakthrough in our understanding of the chemistry of Europa and its relationship with an interior ocean.

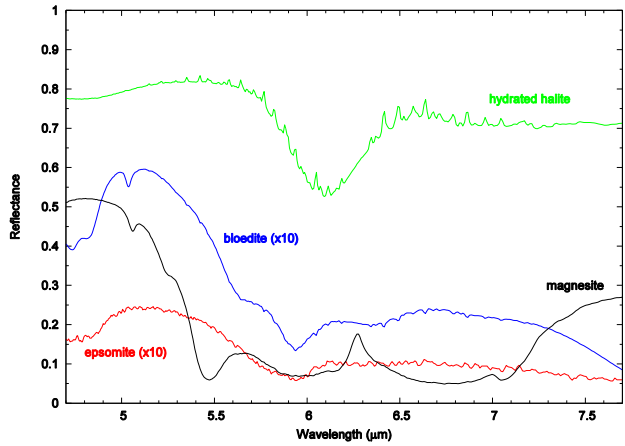

Figure 10. Reflectance spectra of four hydrated minerals proposed for the surface of Europa: bloedite ($Na_2Mg(SO_4)_2 \cdot 4H_2O$), epsomite ($MgSO_4 \cdot 7H_2O$), hydrated halite (NaCl), and magnesite ($MgCO_3$), each with their own distinctive spectral features. Sulfuric acid, in contrast, is featureless at these wavelengths. The spectra of bloedite and epsomite have been scaled by a factor of 10 for visibility. Data from the ASTER Spectral Library (speclib.jpl.nasa.gov).

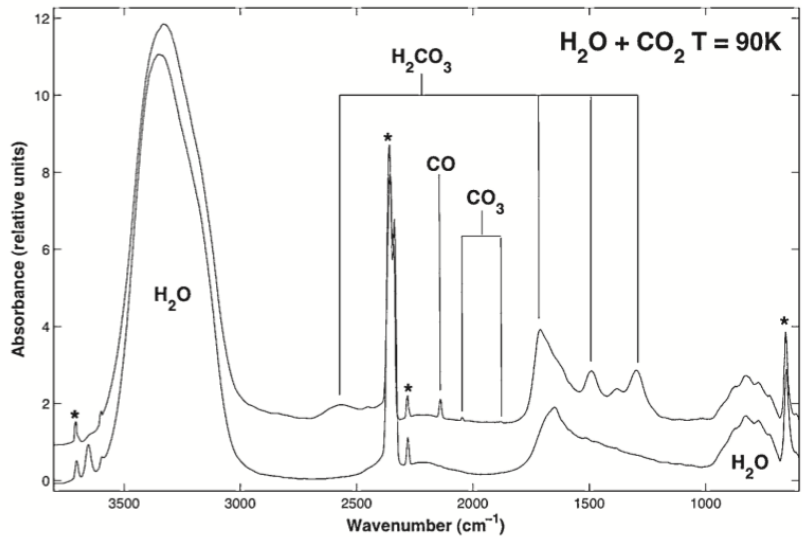

Figure 11. Spectra of $H_2O$ and $CO_2$ ice before (bottom) and after (top) irradiation. Production of carbonic acid ($H_2CO_3$) can be detected by the strong absorption lines beyond 5 μm (wavenumber smaller than 2000 cm$^{-1}$). Such features on Europa would be a clear indication of an active radiolytic carbon cycle and the beginning of the formation of more complex hydrocarbons. From Hand et al. (2007).

Titan

Titan is a target of the very highest astrobiological interest in our own solar system. It is the second largest moon in the Solar System, has a nitrogen atmosphere several times denser at its surface than that of the Earth at sea level, and boasts an active equivalent of the Earth's hydrological cycle in which two hydrocarbons—methane and ethane—take the place of water, each cycling on different timescales (Lunine and Lorenz, 2009; Aharonson et al. 2009).

In its studies of Titan, the US-European mission Cassini-Huygens revealed a world richly endowed in organic molecules on the surface; a series of lakes and seas at polar latitudes hold more hydrocarbon material than the known oil reserves on the Earth (Lorenz et al., 2008). Its nitrogen-methane atmosphere, by virtue of continuous loss of hydrogen liberated from photolyzed methane in the upper atmosphere, is not strongly reducing, and hence is comparable to the pre-biotic Earth's atmosphere in net redox propensity for synthesizing organic polymers. A variety of higher carbon-number hydrocarbons and nitriles are created in the extraordinarily cold atmosphere (94 K at the equatorial surface, 90 K at the poles), which ultimately condense and fall to the surface, where the chemical processes are less well-studied. Some of this material falls directly into the lakes and seas or is transported there by winds; other aerosols agglomerate to form sand-sized particles comprising the equatorial dunes. Hints of variations in the organic composition can be seen in the medium-resolution ($R \sim 150$ at 2 µm) near-infrared spectroscopy performed by Cassini. Ice melted by impacts or volcanism may provide sites where organics react with water over significant timescales to produce amino and carboxylic acids, as well as other precursors to biomolecules.

However, after the final observations of Titan by Cassini are conducted in 2017, gaps will remain in our knowledge due to limitations of the mission. First, geometric constraints associated with the fixed-pallet placement of instruments on the Cassini orbiter (a compromise due to cost) dictate that each of the close flybys will be devoted to only a subset of the instrument techniques. In the end, radar will cover only 40% of the surface at its best spatial resolutions of hundreds of meters. Second, the VIMS near-infrared spectrometer has a lower spectral resolution and sensitivity compared to JWST. The limited spectral resolution is particularly frustrating because Titan's atmosphere must be viewed through a scattering haze of photochemical aerosols and at wavelengths in between the deep absorbing atmospheric methane bands. The VIMS wavelength bands are such that residual methane absorption remains a problem, and atmospheric models must be used to remove this residuum. Finally, there was no possibility for Cassini to cover a full Titan year, 29.5 Earth years. Because of the axial tilt of Titan (essentially coaligned with the spin axis of Saturn), Titan experiences seasonal shifts of sunlight similar in amplitude to that of the Earth. Spacecraft missions to date, and those planned, will cover a portion of Titan's year corresponding to northern late fall through the first "day" of northern summer. JWST's period of operation is unique in that it will cover the portion of Titan's year corresponding to all but the very earliest part of the northern summer.

HST and adaptive-optics ground-based telescopes have achieved diffraction-limited imaging of Titan from Earth. HST NICMOS observations of Titan demonstrated spatial resolution of roughly 200-300 km resolution near the Titan equator and sufficient signal-to-noise to identify the darkest areas as having near-infrared albedos consistent with hydrocarbons (Meier et al., 2000). Ground-based telescopic studies can do what Cassini cannot: provide frequent if not continuous coverage of changes in the atmosphere and on the surface. Ground-based observations suggested short-term changes in the 1-2 µm region of the spectrum associated with the formation of clouds even prior to Cassini (Griffith et al., 2000). In 2008 a major outburst of mid-latitude clouds was observed from the IRTF (Schaller et al., 2009), and in 2010 Cassini observed a major outburst of equatorial clouds, with the surface darkening for weeks thereafter interpreted to be soil damp from methane rain (Turtle et al. 2011).

By 2017 we will be left with the following questions:

- What does the surface look like in higher-resolution ($R \sim 3000$ vs. 200) near-IR spectroscopy?

- What time-variable phenomena might occur due to seasonal (decadal) variations or stochastic surface events in the near-infrared and in that part of the mid-infrared (640 cm$^{-1}$) where the atmosphere is once again optically thin enough to see the surface?

- What are the responses of the troposphere and stratosphere to seasons not observed by Cassini, namely late northern summer and northern autumn (Figure 31)?

- As the intertropical convergence zone crosses the equator, will we see an outbreak of clouds at northern autumnal equinox, as we did at spring equinox?

In the post-Cassini era, the only facilities able to study Titan in the infrared will be JWST, and sites such as Keck and the VLT that have already been observing Titan for years (and not in the mid-IR). JWST can make NIRCam images, and NIRSpec IFU spectral imaging of Titan to build on the 2004-2017 Cassini mission survey, creating a potentially long (10 year +) baseline of spaceborne near-infrared observations of Titan's surface and atmosphere during a seasonal configuration hitherto unexplored in the infrared Figure 12). Of interest is whether surface changes or secular atmospheric changes are in evidence over a decadal timescale. The pixel size on NIRCam gives about the same spatial resolution on Titan as Hubble (Table 2), but the signal-to-noise is much higher. Spectral resolution many times better than that obtained by Cassini VIMS (352 channels from 0.3 to 5.1 µm) can be accomplished using MIRI, over the full range (4.8-5.6 µm) of Titan's 5-µm window, addressing questions of surface composition Cassini cannot answer. Clark et al. (2010) show that with spectral resolution R = 500 and wavelength coverage out to 5.6 µm, key organic molecules thought to be present on Titan's surface can be distinguished with MIRI on JWST that cannot now be detected or distinguished from each other with Cassini. With JWST's instruments, the ability to probe the atmosphere over several levels down to and including the surface over a broad spectral range provides a unique long-term capability.

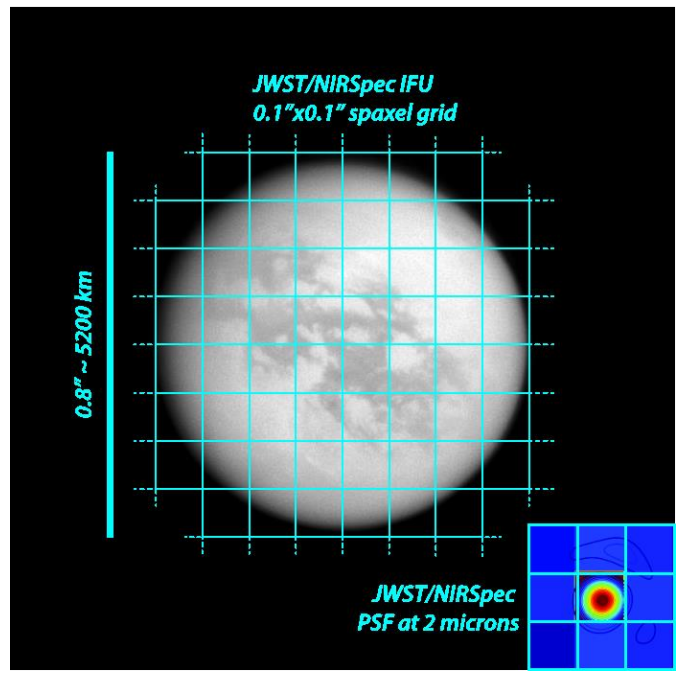

Figure 12. Example of the spatial coverage provided by the NIRSpec IFU on Titan. The Titan image shown here is 2009 Cassini data at 0.94 µm.

One approach is to take NIRCam, NIRSpec and MIRI data on Titan over three equally spaced intervals during the 16-day orbit of Titan, which is phase-locked to Saturn. This provides images and spectra centered approximately 120° apart from each other, and hence global coverage. Cloud movement in the stratosphere, based on our understanding of Titan's winds, occurs with a velocity of 100 m/sec and hence it requires many hours to track cloud movement in detail. Cloud tracking with HST was very difficult because of the telescope's 90-minute orbit around the Earth; with JWST at L2 it will be much easier. Beyond the initial cloud campaign, revisits to Titan once per year should be done to map in imaging and spectroscopy each of the hemispheres to look for longer-term changes.

Changes on the surface or in the atmosphere may present themselves over the next decades. The year 2009 corresponded to the onset of northern hemisphere spring equinox, and the polar regions where large numbers of lakes and one Caspian-sized sea are present experienced sunlight for the first time in almost 15 years. Dramatic changes are expected in the atmosphere above the lakes region and on the surface itself. JWST will be able to monitor these changes beginning a year after the end of the Cassini mission in 2017.

Other Icy Satellites

The satellite systems of the giant planets are host to great diversity as well as insight into such topics as volatile transport, low-temperature chemistry, and the history of the Solar System. With JWST, these bodies may be studied with unprecedented detail. The satellites' surfaces display prominent near-infrared ice signatures that NIRSpec can observe with high spectral resolution. In the mid-infrared, the thermal properties of these objects may be studied with MIRI observations with greater signal-to-noise than Spitzer. The superior collecting area of JWST will also facilitate observations of smaller objects, enabling near- and mid-infrared spectroscopy of most outer irregular satellites for the first time.

Examples of icy-satellite projects possible with JWST observations include

- Investigation of longitudinal (and possibly latitudinal) variations in surface composition, particularly ices. This may benefit models of inter-satellite relations (such as sputtering from Phoebe's surface being related to Iapetus' varied surface [Tamayo et al. 2011]) and models of volatile transport on individual satellites. Seasonal effects may also be determined.

- Characterization of surface ices to determine their physical state, relative abundances (e.g., $CO_2$ vs. $H_2O$), and chemical evolution. For example, the properties of the surface ices may address the degree to which irradiation plays a role in their physical/chemical state, as seen in the "Pac-Man" features on Mimas and Tethys (Howett et al. 2012).

- Analysis of the chemical composition of tiny satellites to determine their origin, particularly through comparison with ring material and comets. Any identifiable differences among the satellites may indicate contrasting histories.

- Assessing satellites' mid-infrared spectral energy distributions to constrain their internal heat levels, which will benefit models of orbital migration. Such observations may also determine whether Miranda and Ariel are active (or were in the recent past) (Pappalardo and Schubert 2013; Castillo-Rogez and Turtle 2012), and whether Triton may support a subsurface ocean (Gaeman et al. 2012).

While the Cassini mission has explored the Saturnian satellites in the near- and mid-infrared with high spatial resolution, there are ways JWST observations can improve this dataset. NIRSpec offers a greater spectral resolution ($R = 2700$) compared to VIMS ($R \sim 200$), allowing better identification and characterization of key absorption features. JWST will also fill the gap between 5 and 7 μm not covered by VIMS and CIRS. Furthermore, JWST will extend the temporal baseline of these satellite observations, extending the seasonal coverage of the Saturnian system through the approach to southern solstice.

For most satellites, observations at the four cardinal points in the orbit are desired to obtain coverage of the sub-planet, anti-planet, leading, and trailing hemispheres. For most principal satellites, it is possible to acquire spectra at all four positions within a single epoch; the only such satellites with orbital periods longer than ~20 days are Iapetus (79 days) and Phoebe (1.5 years). Seasonal considerations will be important in planning such observations as different latitudes enter or leave view. This will be particularly important for the satellites of Uranus and Neptune due to their long seasonal cycles. Certain latitudes of Uranian satellites will only be visible at the start of the JWST mission (as Uranus will be approaching its 2030 solstice), while certain latitudes in the Neptunian system will only be visible at the end of the mission (in the approach to equinox in 2038). Effects due to seasonal changes in the insolation pattern may be monitored as well, particularly for active bodies like Enceladus and Triton, and for satellites whose activity is unknown (such as Ariel).

Multiple-Satellite Spectroscopy

In addition to standard individual observations of satellites, near-simultaneous spectroscopy of multiple satellites may be possible with NIRSpec's microshutter assembly (MSA). While tracking on a background star, the central planet is held fixed in the NIRSpec field of view for an extended period of time. Groups of shutters would be held open in anticipation of satellites passing through them during this time period, as shown in Figure 13. Spectra from multiple adjacent slits may be combined in cases where a satellite crosses multiple apertures, and/or where the satellite in question is larger than the size of an aperture. With this method, one hour of observation may result in numerous several-minute spectra totaling ~1 hour or more. This scenario presents an alternative to the standard method of observing satellites one by one: less time is required for system procedures, though at the cost of the satellite spectra accumulating background signal over the full duration. (Note that data processing associated with these observations will have to take into account the fact that the satellites will not be present in their apertures for the full integration.)

This method is best suited for observing the inner satellites of Uranus and Neptune, due to their faster orbital motion and the lack of a bright ring system (although issues with Saturn's rings will be minimized during the planet's 2025 equinox). Outer large satellites, having slower angular speed, tend to linger in the same aperture long enough to saturate; keeping the background stars fixed while allowing the planet and satellites to move through the field may counter this effect, at the expense of losing inner-satellite spectra due to interference from the planet itself or other satellites. Outer irregular satellites tend to be spaced farther apart than is feasible for NIRSpec's field of view.

At present, this method is not supported by MSA capabilities, as MSA target acquisition is not currently set up to coordinate with moving target observations. If the tracking ability is expanded to facilitate holding a moving target fixed within the field of view, then the satellite observation process outlined above may become a viable option.

Satellite Discovery and Astrometry

Observations with JWST will be particularly useful in efforts to discover new satellites. While the available field of view is too small for initial investigations, the telescope's large collecting area will be particularly valuable in follow-up observations of potential targets identified by large-scale surveys. Since most outer irregular satellites are expected to have low albedos, observing their mid-infrared thermal radiation is ideal for such observations. However, since the background signal (largely zodiacal light) also becomes brighter in this spectral region, the optimal MIRI filter varies depending on the satellite's expected temperature. Estimates of the minimum satellite sizes detectable with MIRI imaging are given in Table 4.

Among the confirmed satellites, some small satellites have orbits that are chaotic or otherwise not well constrained: particularly Saturn's moons Prometheus and Pandora (Farmer and Goldreich 2006), and Uranus' moon Mab (Kumar et al. 2011). Earth-based monitoring of such satellites has been difficult due to their small sizes. Given JWST's great sensitivity to faint objects, JWST observations will greatly contribute to these satellites' astrometric datasets.

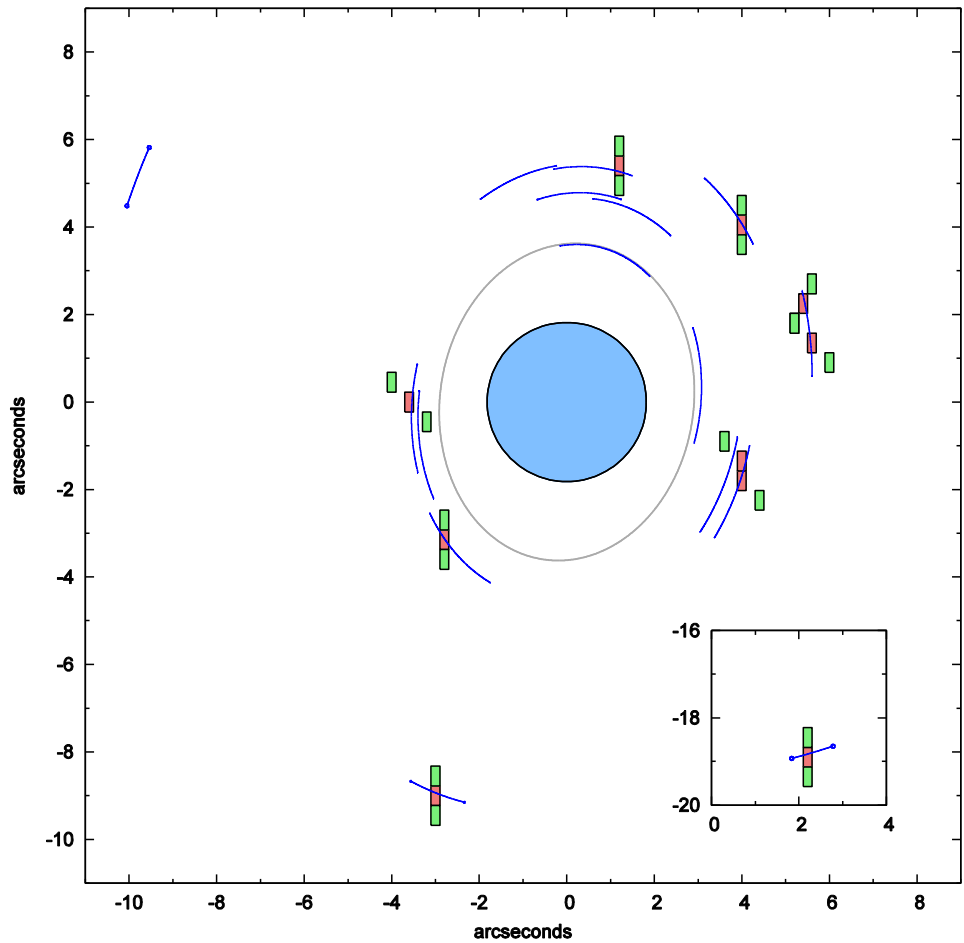

Figure 13. An example MSA configuration for observing the inner Uranian system for 60 minutes, at a randomly chosen time on September 28, 2020 (a valid time for observing Uranus with JWST). Uranus is shown at center, with the Epsilon Ring in gray. Open apertures are shown as colored rectangles: red for apertures that will result in object spectra, and green for nearby "empty" spectra to use for calibration. Observed satellites include (from top to bottom) Cupid, Puck, Mab (with two apertures), Cressida, Perdita (with two apertures), Desdemona, Miranda, and Umbriel (inset). Ariel, at left, is unobservable in this arrangement as it will saturate. The shortest-timespan spectrum will be that of Cupid, at 6.8 minutes. This scenario will produce satellite spectra totaling 2.1 hours.

Rings

The rings that adorn the four giant planets are of prime importance as accessible natural laboratories for disk processes, as clues to the origin and evolution of planetary systems, and as shapers as well as detectors of their planetary environments (Tiscareno, 2013). The retinue of small moons accompanying all known ring systems is intimately connected as both sources and products, as well as shepherds and perturbers, of the rings.

Imaging

In the context of rings, observations of faint targets are complicated by the nearby presence of the bright planet. Strategies are needed to enhance the apparent brightness of desired targets and/or to suppress the apparent brightness of the planet. JWST will be equipped with filters that allow it to image giant planet systems at wavelength bands in which the planet is greatly darkened by atmospheric absorption due to methane and other atmospheric constituents. For observations of faint moons or rings that are close to bright giant planets, this will lead to greatly improved signal-to-noise and spatial resolution comparable to HST and other observatories operating in the same wavelength bands. (Put another way, JWST will operate within the infrared methane bands at a spatial resolution comparable to that at which HST operates in visible bands, with vastly improved signal-to-noise when suppression of glare from the planet is an important factor.) As a result, JWST will provide major advances in resolving and separating the main rings of Uranus and Neptune, improving upon HST and ground-based observations of their fine structure (de Pater et al., 2005, 2006, 2007; Showalter and Lissauer, 2006).

The best wavelength at which to observe the rings of the giant planets would be 2.3 μm, a wavelength at the center of a strong methane absorption feature but where water ice is bright. However, NIRCam does not have a filter centered at 2.3 μm. Desired 2.3-μm imagery may be produced by extracting the relevant spectral region from a NIRSpec IFU cube, though further study is required to determine the feasibility of this method. Imaging alternatives may include NIRCam filters F182M, F187N, F323N, and F335M, which lie mostly within methane absorption features. In the case of Saturn's rings, imaging longward of 2.8 μm will be difficult due to water ice becoming very dark.

JWST will have new sensitivity to yet-undiscovered faint rings, including the predicted rings of Mars (Showalter et al., 2006) and Pluto (Steffl and Stern, 2007). The New Horizons spacecraft, which conducted a 2015 flyby of Pluto, will likely not have the last word on Pluto's possible rings due to its flyby speed and limited range of viewing geometries. JWST will be ideal for follow-up observations, possibly with greater sensitivity, and can also search for rings around other trans-Neptunian dwarf planets.

Continuing to observe and track objects that are faint, recently discovered, or known to be changing is of high importance. JWST observations will be important for continuing to track the evolving ring arcs of Neptune (de Pater et al., 2005), the progressively winding ripple patterns in the rings of Jupiter and Saturn that trace cometary impacts (Hedman et al., 2011; Showalter et al., 2011), and other faint targets.

Spectroscopy

The compositional diversity of solid objects in the outer Solar System is apparent from the near-infrared spectra of bodies such as Triton, Pluto and Charon, which show absorption features of varying strengths due to varying amounts of methane, water and other ices on their surfaces. The smaller moons and rings of Neptune might have originally been made of the same stuff as these larger objects, but they also would have had much different evolutionary histories (perhaps less thermal processing, more pollution from infalling matter, etc.). Comparing the surface composition of these smaller objects to their larger neighbors should therefore help clarify the origins and histories of both, but it is difficult to obtain good-quality spectra of these very small and/or faint objects from ground-based observatories.

With its large mirror and high-quality spectrometer, JWST will be able to take spectra of very faint objects. Potential targets include the rings and small moons of Uranus and Neptune; these have never been the subjects of high-fidelity spectroscopic study, as Voyager 2 did not carry a spectrometer capable of detecting them. Investigators will be able to use JWST observations to measure the strength of water ice features, as well as measure the band strength of features from other possible constituents, such organics (with prominent features near 3.3-3.5 µm and longward) and silicates. Characterizing their chemical compositions is of considerable interest for addressing the origins of the Uranus and Neptune systems as well as for addressing the question of why the Uranian and Neptunian rings are so qualitatively different from those of Saturn (Tiscareno, 2013).

By the same token, JWST will be able to acquire very sensitive spectra of all objects over a broad range of wavelengths. It will be able to fill in the gap between Cassini VIMS and Cassini CIRS (from 5 to 8 µm) and will be able to map Saturn's rings in the 1.65-µm water absorption feature (which falls in an internal gap in VIMS' spectral coverage, and is unusual in that its depth is useful for mapping temperature variations). Its spatial resolution will be comparable to CIRS, and its sensitivity will be greater, so it should be capable of improving current maps of Saturn's rings in the thermal infrared (though over a very limited range of phase angles) and may achieve the first detection of the faint silicate absorption features at 10 µm, yielding information about the little-understood non-water-ice components of Jupiter's and Saturn's rings.

Figures 14 and 15 show the best ring spectra available at low phase angles (as would be the case from JWST). Only Saturn's rings have detailed spectra at such conditions, taken by Cassini VIMS (Hedman et al., 2013). The best spectra taken to date of Uranus' rings were taken by the Keck telescope in Hawaii (de Kleer et al., 2013), but they are very noisy and are ripe for improvement by JWST, especially since they do not go beyond 2.5 µm. Among these spectra the compositional differences between the two ring systems are clear: Saturn's ring spectra are dominated by water-ice absorption features, while the absence of such features in Uranus' rings indicates low water ice content. Spectra of the rings of Jupiter and Neptune are in even greater need of improvement, as no spectra of quality have been taken of either system at low phase angles. Galileo NIMS took spectra of Jupiter's rings at very high phase angles (e.g., Brooks et al., 2004), but these are dominated by diffraction and do not indicate the spectral features that JWST would see.

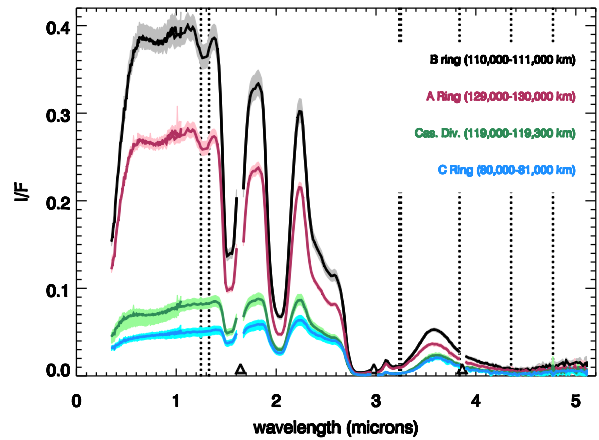

Figure 14. Average Cassini VIMS spectra of the lit face of selected regions in Saturn's main rings at low phase angles. All of the absorption features present are due to water ice; other ring constituents are either featureless in the near infrared, or not present in large amounts. Figure from Hedman et al. (2013).

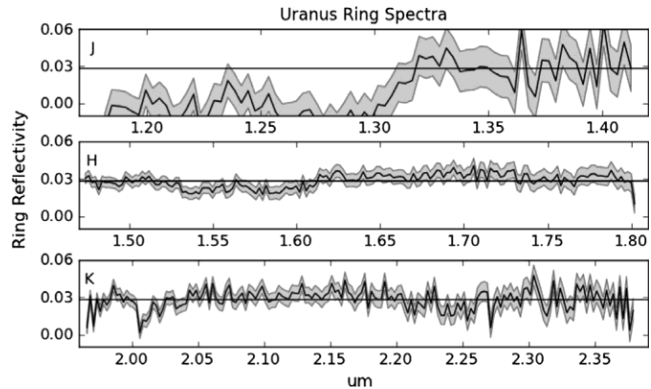

Figure 15. Keck spectra of Uranus' main rings. Note the lack of prominent water-ice features, when compared to the spectra of Saturn's rings. The very low albedo suggests these rings are carbon-rich. Figure from de Kleer et al. (2013).

Observing Geometry

The next Saturn equinox will take place in 2025. The event itself will not be observable by JWST, as it will occur when Saturn is near the Sun as seen from Earth, but low sun angles will be observable approximately three months before and after equinox. This will facilitate the observation of seasonal phenomena such as spokes, which are prevalent near equinox and absent near solstice (Mitchell et al., 2006, 2013). JWST will have sufficient resolution to continue monitoring spokes, as has HST (McGhee et al., 2005), which will have particular value after the end of the Cassini mission. JWST will also be able to improve on the tracking of clumps in and around the F Ring near equinox (McGhee et al., 2001), and will enjoy optimal edge-on viewing of Saturn's dusty rings during this season (de Pater et al., 1996, 2004). (The Phoebe ring, which lies in Saturn's orbit plane and is always edge-on as seen from Earth, is thus always available for optimal edge-on viewing.)

Neither Uranus nor Neptune has an equinox that falls within the JWST mission (Figure 16). During the JWST mission, Sun angles will decrease at Neptune (solstice 1997, equinox 2038), and will increase at Uranus (equinox 2007, solstice 2030). This will lead to increasingly favorable viewing for both ring systems as the JWST mission progresses, since Neptune's rings are primarily dusty while Uranus' rings are dense and sharp-edged. The only exact equinoxes possibly observable by JWST will be at Jupiter; these will provide optimal viewing of vertical structure in the halo/gossamer rings.

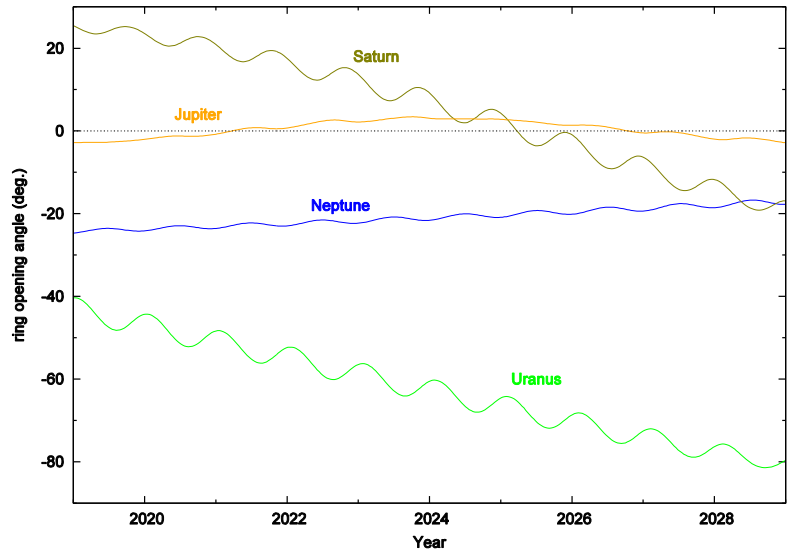

Figure 16. Opening angle as a function of time for known ring systems from 2019 to 2029 (Meeus 1997).

Primitive Bodies

The small objects in the Solar System (e.g., asteroids, comets, and trans-Neptunian objects) will be of particular interest for study with JWST. This follows because they are thought to be 'primitive', preserving information about processes and chemistry in the solar nebula, and from the fact that many of them are too faint to have been studied in detail from the ground at near- to mid-IR wavelengths. The sensitivity of JWST will allow for compositional characterization of objects much smaller and further away, and with very high data quality. The modest spectral resolution of the instruments is well matched to studying the solid surfaces of these airless bodies, and will be useful for performing compositional studies of the gas comae of comets. These statements also apply at mid-IR wavelengths, but MIRI will additionally allow for the determination of albedos, diameters and thermal properties of a wide range of objects previously too faint to have such measurements. The trojan asteroids, centaur objects and irregular satellites of the planets also fall into this category.

Comets

Comets provide important clues to the processes that occurred during the formation and early evolution of the Solar System. In addition to their significant inventories of volatiles, they have orbits with aphelia in the outer Solar System, which are unstable due to perturbations by the giant planets. These properties suggest that comets are objects that have resided in the outer Solar System since the time of planetesimal formation and have only recently been perturbed onto orbits entering the inner Solar System. The degree of devolatilization and thermal processing must vary between comets, but as a class they represent packages of relatively pristine material which are uniquely suited to study because they come close to the Earth and Sun, and their outgassing provides the opportunity to determine the composition of materials in their interiors. Past observations, as well as laboratory measurements of cometary material obtained from the Stardust mission, suggest that comets contain a mixture of both interstellar and nebular material. A major observational challenge in cometary science is to quantify the extent to which individual comets, or parts of comets, can be linked to either reservoir.

JWST will support studies of many aspects of comets and their behavior, including being able to track even relatively fast-moving targets. Imaging will provide information on nuclear composition, diameter, albedo and thermo-physical properties. Regarding the dust coma, imaging will also enable investigation of the production rate and distribution; jet activity; activity evolution; and the dust temperature, albedo, and emissivity. Because JWST will offer such a huge improvement in spatial resolution in the mid-IR, near-nucleus processes in the dust coma may be particularly amenable to study via imaging. Imaging will also provide some compositional information regarding cometary nuclei and dust, particularly for small and/or distant objects and those with very small dust production rates. The available filters should allow discrimination of various silicate components and the identification of water and other ices.

JWST will support compositional studies of cometary nuclei, gas and dust with unprecedented sensitivity throughout the 1-30 μm range via the various spectroscopic modes. These data will allow determination of the composition, grain-size and temperature of dust in the coma and of the nucleus. The spectral resolution of the JWST instruments is not high enough to support all studies of cometary gases that can be done from the ground, but will provide very high sensitivity and spatial resolution, as well as access to wavelengths that can't be observed from the ground.

The Spitzer IRS instrument spanned the MIRI wavelengths, and provided unprecedented (c. 2005) sensitivity. Figure 17 is the thermal emission spectrum from the coma of 29P/Schwassman-Wachman 1 (29P/S-W 1) taken using IRS when the comet was at a heliocentric distance of about 6 AU. These can be compared to the sensitivity of MIRI, illustrated in Figure 18 and Figure 19. The IRS data and Spitzer MIPS 24um imaging data (Stansberry et al. 2004) showed that the nucleus of 29P/S-W 1 has a diameter of about 25 km; the coma optical depth was estimated as $10^{-9}$, with a production rate less than 50 kg/s. MIRI will be capable of measuring much weaker activity, via imaging and imaging spectroscopy, and at higher spectral resolution than IRS. MIRI spectra will be capable of providing completely new levels of detail regarding mid-IR emission from silicate grains and molecules active in the mid-IR, and for small and/or distant and/or weakly-active comets that simply couldn't be studied at all using earlier facilities. The MIRI imager sensitivity to extended emission from cometary comae and trails is also good, enabling high spatial-resolution maps at dust optical depths as low as $10^{-5}$ at 6 AU using exposure times of only about 100 seconds. More tenuous dust, or dust around more distant comets, could be mapped at high SNR using longer exposures and/or by binning MIRI pixels.

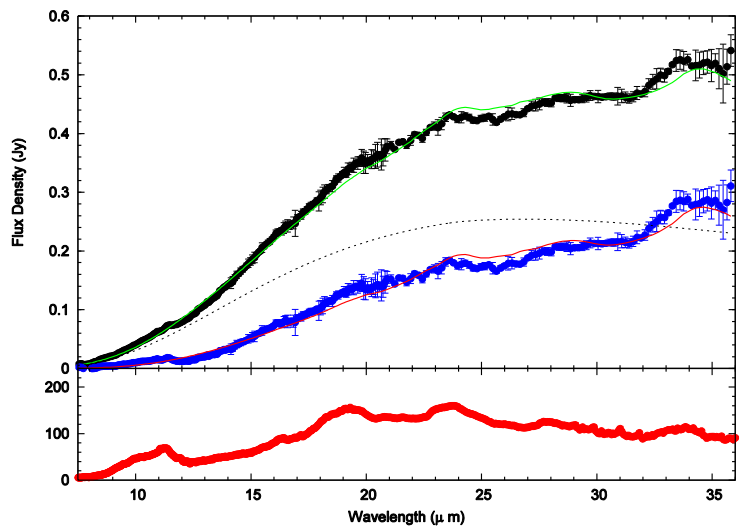

Figure 17. The emission spectrum of comet 29P/Schwassmann-Wachmann 1 (top panel) made using the IRS instrument on Spitzer Space Telescope (from Stansberry et al., 2004) compared to the ISO spectrum of comet Hale-Bopp (lower panel; Crovisier et al. 1997). The 29P spectra are as measured (black points) and after removal (blue points) of a model of the thermal emission from the nucleus (black dotted line). Note that 29P was more than 100 times fainter than Hale-Bopp.

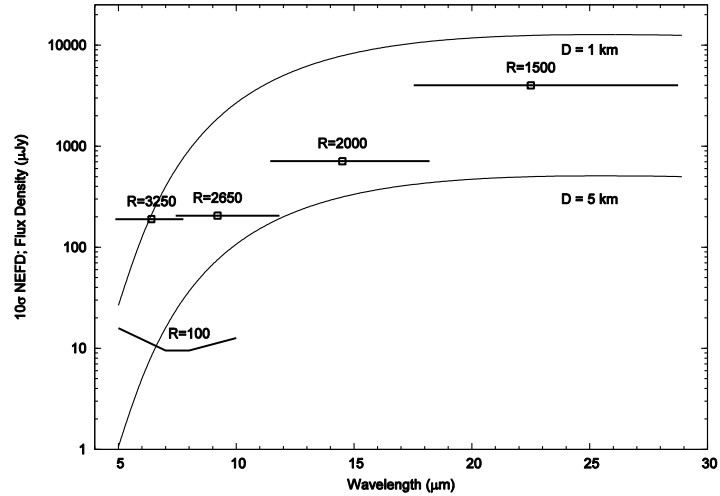

Figure 18. The 1000-second continuum sensitivity of MIRI low-resolution spectroscopy compared to models of the thermal emission from comet nuclei at 3 AU, where coma emission is typically negligible, with $p_V = 5\%$.

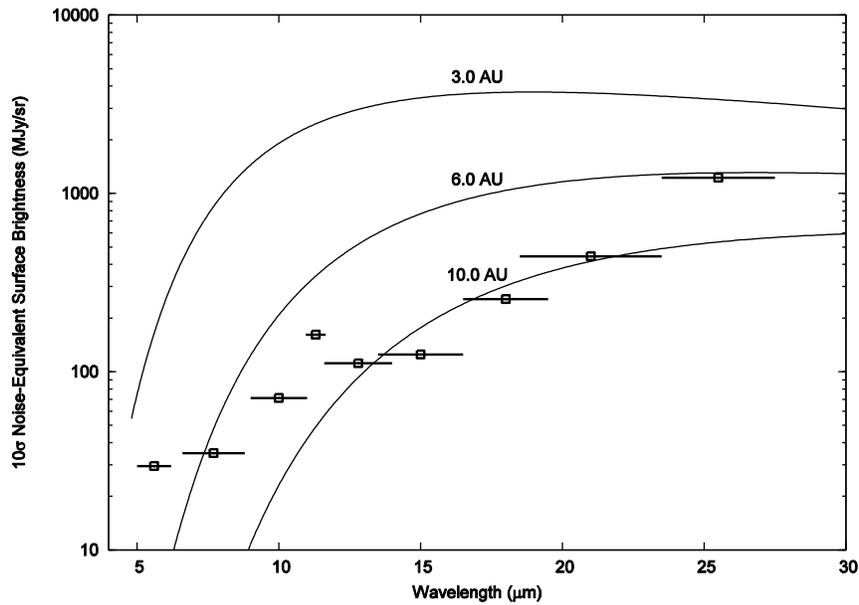

Figure 19. MIRI per-pixel 1000-second sensitivity to extended emissions is compared to models of emission from coma dust at a range of distances from the Sun. The dust optical depth is assumed to be $10^{-5}$ in all cases. MIRI will provide outstanding capabilities for detecting and mapping the dusty comae of relatively active comets at high spatial resolution. Pixels may have to be binned to detect this emission, resulting in lower spatial resolution but high sensitivity.

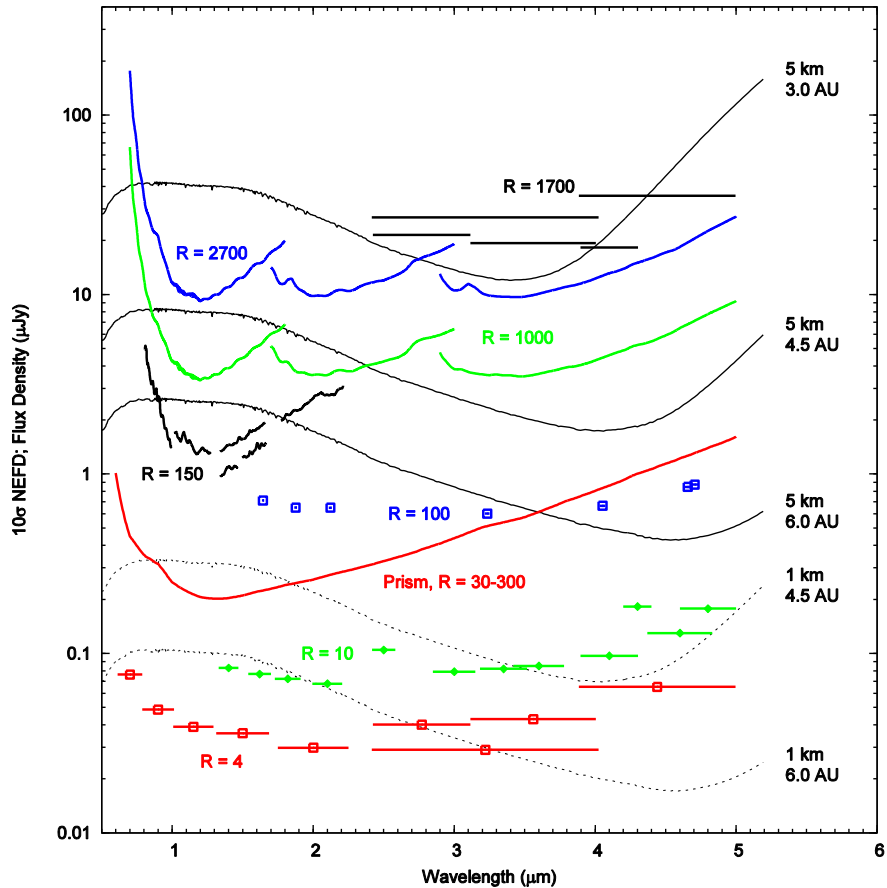

Figure 20. JWST sensitivity (thick lines/curves) compared to models of reflected plus emitted spectra for comet nuclei (thin curves). The sensitivity curves noise-equivalent flux density, or NEFD) are for a 1000-second exposure and a signal-to-noise ratio of 10. Thick colored curves are for NIRSpec at spectral resolving powers $30 < R < 300$ (prism), 1000, and 2700. Thick black curves are for the NIRISS $R$=150 slitless grism (paired with several different bandpass filters). The gold horizontal lines are for the NIRCam $R$=1700 slitless grism, also paired with available bandpass filters. The other horizontal lines (all below 1 µJy) are NIRCam sensitivity limits. The thin black curves are the model spectral distributions for comet nuclei at different distances from the Sun and JWST (assumed equal). Objects are assumed to be spectrally neutral with a visual geometric albedo of 5%, and diameters of either 5 km (solid curve) or 1 km (dotted curve), making them effectively point sources.

MIRI Low-Resolution Spectrometer (LRS) data from 5-14 µm will be sensitive to the emission features of silicates, PAHs and other large organic molecules. The LRS data also provides sensitive constraints on the temperature of the dust, and grain size distribution. The LRS data will only be taken on the near-nucleus region of the coma. MIRI can also be employed to collect 10-µm images of the coma and near-nucleus dust trail. The images will reveal jet structures in the coma, providing constraints on the rotation state of the nucleus; the dust production rate, and the velocity of ejection. Repeat observations may be performed to sample activity at different phases in a comet's orbit.

NIRSpec will be capable of detailed characterization of the composition of comet nuclei, as shown in Figure 20. The NIRCam and NIRISS grism modes could also be useful for studying small, distant comets, particularly if they are too faint, or have ephemeris uncertainties too large, to allow target acquisition to hit a NIRSpec slit.

NIRSpec medium-resolution spectra will enable characterization of broad emission lines from $H_2O$, $CO_2$ and organic molecules in the gas phase (Figure 21). These spectra will also be highly sensitive to the presence of water ice and silicates in the dust grains of the coma. Spectra will be taken of the region surrounding the comet nucleus to characterize the gas and dust composition before interactions with UV and chemical evolution have taken place, and also at a position offset from the nucleus to characterize the photo/chemical processes in the coma. NIRSpec high resolution spectra will be used to measure abundances of higher-order organic molecules, as well as isotope ratios of some volatiles in comae.

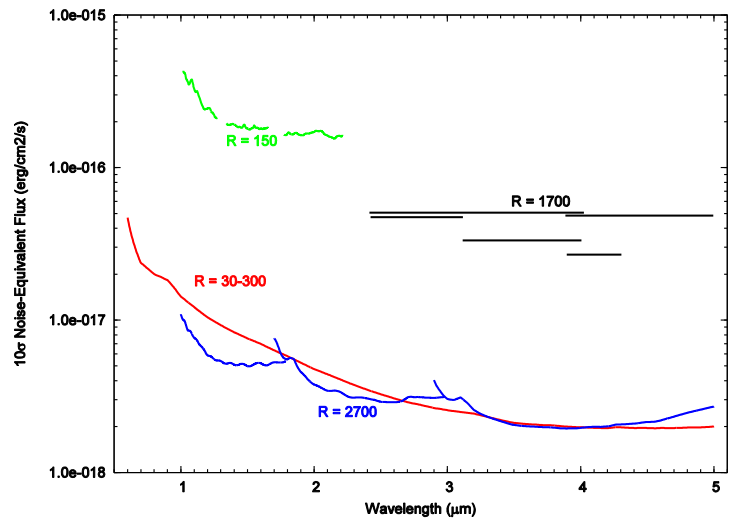

Figure 21. The sensitivity of NIRSpec to emission in unresolved lines, for 1000-second exposures. At resolving power 2700, NIRSpec can detect and differentiate emissions from many cometary molecules, including $H_2O$, HDO, $CH_4$, CO, OH*, HCN, $C_2H_6$ and $CH_3OH$, using modest exposure times. The NIRCam and NIRISS grisms are much less sensitive to such emission due to their lack of slits.

NIRCam and NIRISS could be used for characterizing the composition of faint comet nuclei via multi-filter imaging. Depending on the heliocentric distance, some of the emission could be thermal, complicating interpretation of such data. Comets at distances beyond 6 AU should have almost no thermal emission shortward of 5 µm, and these instruments are sensitive enough to measure the near-IR colors of objects as small as 1 km using relatively short exposures.

Kuiper Belt Objects

JWST offers the possibility of vastly improving our understanding of the composition of objects in the trans-Neptunian region. These bodies, known as TNOs or Kuiper Belt Objects (KBOs), are thought to be planetesimals formed in the proto-solar disk and nebula and which never accreted to form planets. The KBO population does, however, include all but one of the five dwarf planets: Pluto, Eris, Haumea, and Makemake. The Kuiper belt and

KBOs are in many ways similar to the asteroid belt and asteroids. Dynamically, the asteroid belt is influenced by various resonances with Jupiter, while the Kuiper belt was sculpted by the outward migration of Neptune. The asteroid belt contains many 'families' of objects, which formed via disruptive collisions; the Kuiper belt contains only one known collisional family, so far. On the other hand, the Kuiper belt does contain several dynamically distinct classes of objects that are not collisional families: the classical KBOs, with dynamically cold and dynamically hot subclasses (which differ by having low or high values of inclination); the resonant KBOs (of which Pluto is the archetype), the scattered KBOs (e.g., Eris and Sedna); and the centaurs, KBOs that have been recently perturbed onto non-resonant orbits with perihelia inside Neptune. The provenance and history of these dynamical classes are one of the key questions regarding the Kuiper belt.

The primary difference between the two populations is compositional, with the asteroids being dominated by silicates with significant admixtures of organics and metals. The KBOs are much more volatile rich, with large fractions of water but also other species such as nitrogen, methane, carbon monoxide and dioxide, although they too must contain significant amounts of silicates. The largest KBOs have densities of about 2 g/cm$^3$, consistent with about a 50:50 mix of silicates and water. Interestingly, the fifth dwarf planet, Ceres, which resides in the asteroid belt, has the same density as the KBO dwarf planets, suggesting that it is probably very water rich. The KBOs are thought to be volatile rich because they formed in the outer solar nebula and disk where such species were cold enough to condense. Because of this, studying the composition of KBOs should provide unique insights into both the volatile content of the early solar nebula, and into processes that operated in the outer parts of the nebula and disk during planet formation. The volatile species present on KBOs, as well as many organic compounds known or expected to exist on their surfaces, also happen to have their vibrational absorption bands in the near-IR where they are amenable to characterization with NIRCam, NIRSpec and NIRISS. Many silicate absorption/emission features are also present, some in the near-IR but also throughout the MIRI wavelengths. MIRI also will allow the determination of surface temperature distributions on KBOs, via modeling of the slope of their SEDs. This information is diagnostic of the thermal inertia and roughness of the surfaces, providing additional insights into the regolith properties of these objects. Those properties are, in turn, probably influenced by the collisional environment in the trans-Neptunian region.

The figures below illustrate the sensitivity of the JWST instruments in comparison to the reflected and emitted radiation from KBOs. If JWST were to take an extended series of short exposures (<100 seconds, to minimize KBO travel from one pixel to another) and they were to be combined to account for the expected rate of motion for KBOs, 50-AU objects as small as 20 km ($m_V$ ~ 28.6) could be detected with the F150W filter using 5000 seconds of exposure time. However, the narrow field of view would make this detection method inefficient under most circumstances.

Due to limits on the sensitivity of the ground-based instruments and observatories used to find KBOs and determine their orbits, the smallest known KBOs are probably about 50 km in diameter. A vast majority of the known population have diameters of 100 km or larger. JWST and its instruments offer such a quantum leap in sensitivity that it would be possible, in principle, to obtain high-quality near-IR multiband photometry for every known KBO. Additionally, many fundamental absorption features occur in the L and M bands, which are very difficult to observe at adequate sensitivity from the ground. Using NIRCam it will be possible to obtain near-IR colors for a statistically significant sample of objects from each of the dynamical classes mentioned earlier, as illustrated in Figure 22. Because most KBOs are closer than 45 AU from the Sun (especially the smaller ones), the example shown is somewhat conservative. Such a data set would expand the existing near-IR colors for large TNOs and could form the basis of a spectral classification scheme, similar to that developed for asteroids, for the Kuiper Belt.

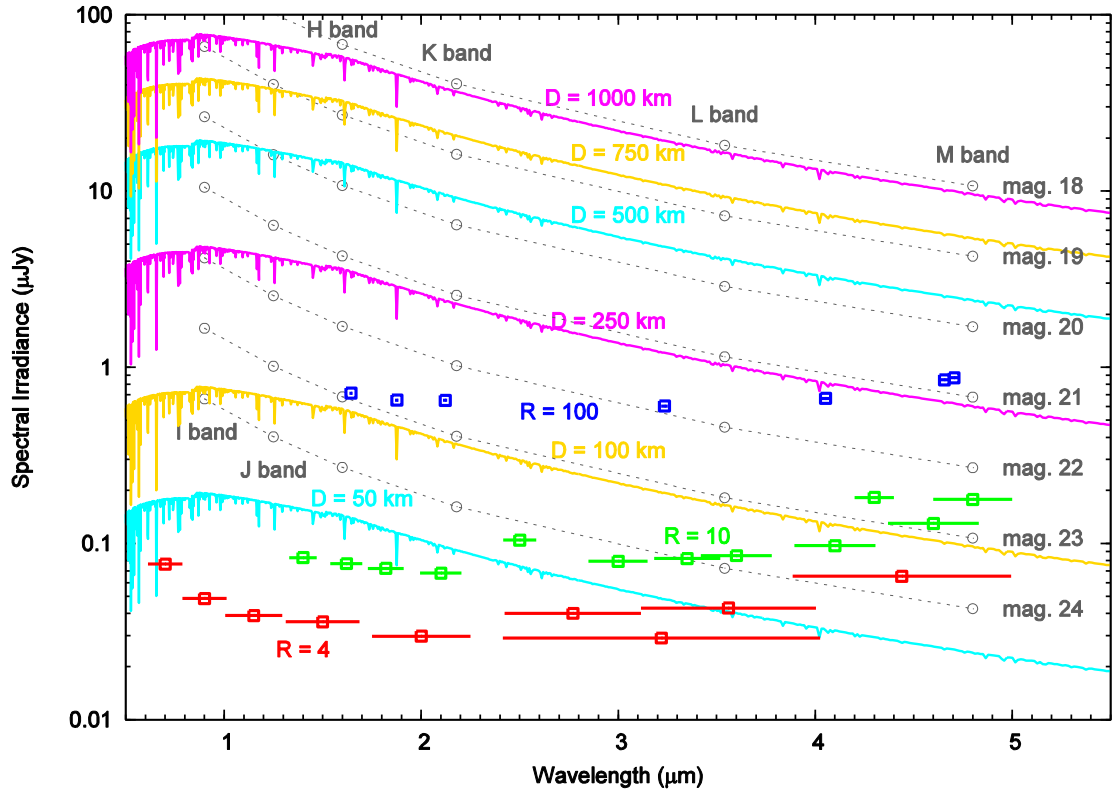

Figure 22. Comparison of model KBO spectra to observed magnitudes in various bands, and to NIRCam 1000-second point source sensitivities. KBO spectra were generated using a Phoenix spectral model (Husser et al. 2013) at a resolution of 360. The KBOs are presented at various diameters, assuming a distance of 45 AU from both the Sun and from JWST, and gray geometric albedos of 0.10.

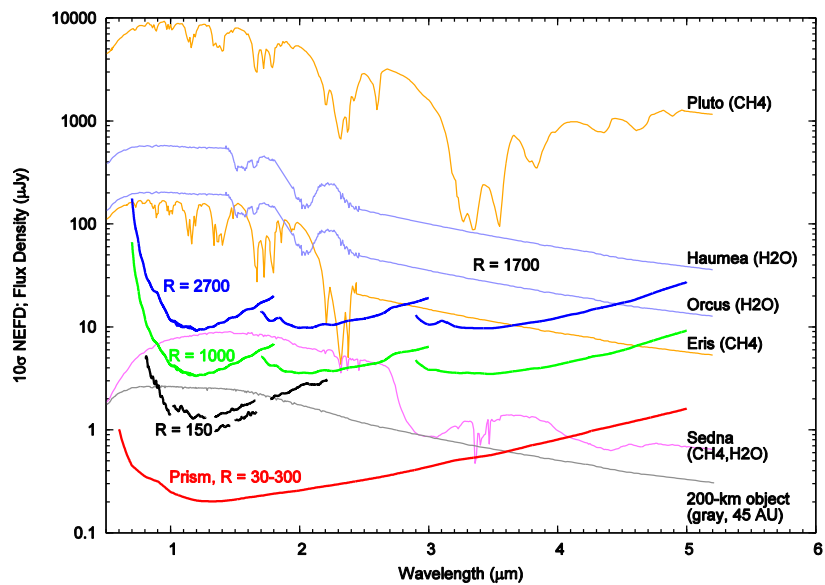

Figure 23. JWST spectral sensitivity (noise-equivalent flux density, or NEFD) curves compared to models of reflected plus emitted spectra for Kuiper Belt objects. The sensitivity curves are for a 1000-second exposure and a signal-to-noise ratio of 10. Thick solid lines are for NIRSpec at spectral resolving powers $30 < R < 300$ (prism), 1000, and 2700. Thick dashed lines are for the NIRISS $R$=150 slitless grism (paired with several different bandpass filters) and for the slitless NIRCam $R$=1700 slitless grism, also paired with available bandpass filters. The thin curves are the model spectral distributions for five named Kuiper Belt objects, with the dominant ice causing the absorptions noted for each. Other ices identified by weaker absorption features on KBOs are due to CO, $CO_2$, $N_2$, $NH_3$, and methanol ($CH_3OH$). The spectrum shown for Orcus is a scaled version of the Haumea spectrum. Spectra were furnished by D. Cruikshank (Pluto), F. Merlin (Eris), J. Emery (Sedna), and K. Barkume (Haumea). Also included is a hypothetical spectrally neutral ($p_V$=0.1) object 200 km in diameter, located at 45 AU. Model spectra were computed using observing circumstances estimated for mid-2019.

NIRSpec will enable high-SNR spectral studies of a significant sample of KBOs, as illustrated in Figure 23. Using the prism, spectra could be obtained for even very faint/small objects, and would be highly diagnostic for water ice and some of the volatile ices (although only $CO_2$ would likely be retained on objects with diameters less than about 1000 km).

Using simple simulations of relatively short observations (three 970-s exposures) of KBOs with the NIRSpec IFU and at low spectral resolution (from 30 to 300 across the 0.6-5.0 µm range), we have computed a first order estimate of which fraction of water ice could be unambiguously detected as a function of the diameter of the object. The results are shown in Figure 24. Even for small objects (diameters as small as 400-500 km), it is possible to detect the signature of water ice with a high level of confidence even for (geometrical) dilution factors of 10 (i.e. fraction of 10%). For reference, in our simple model, an object with a diameter of 500 km would have an H-band magnitude of typically 19-20 (Figure 22).

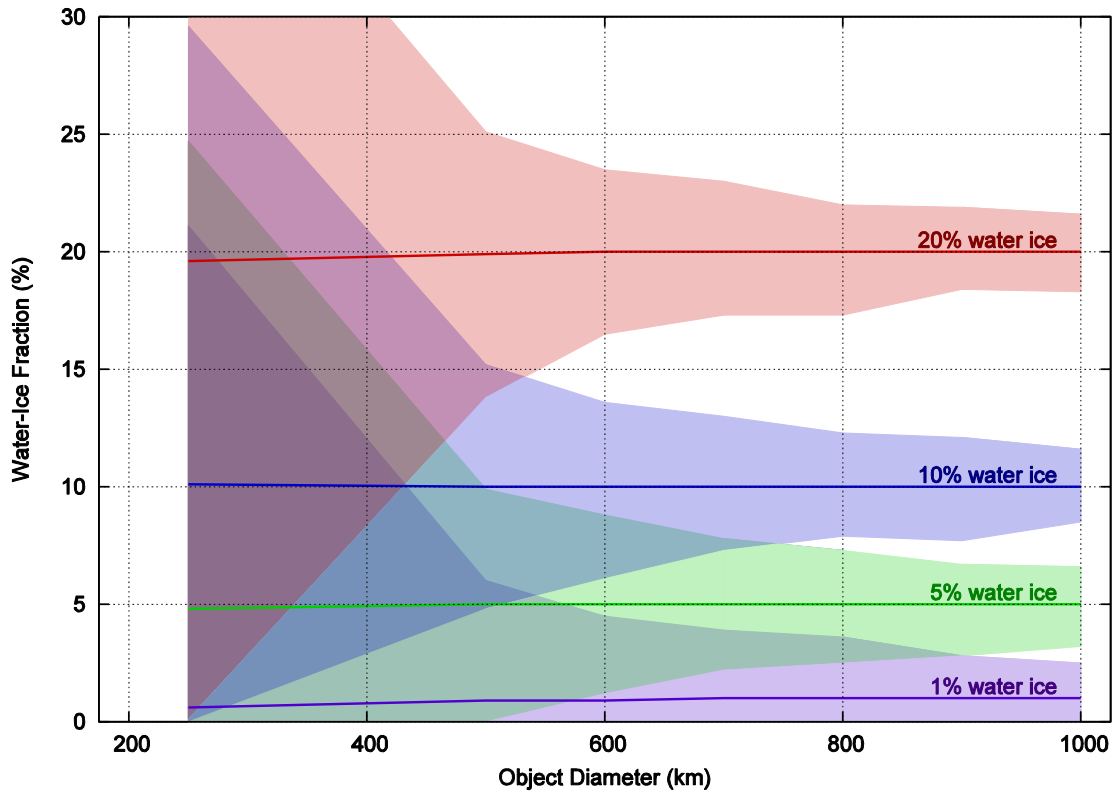

Figure 24. Simulated estimates of KBO water ice fractions derived from NIRSpec IFU observations. The accuracy and precision of water ice fraction estimates improve for larger objects. Objects are modeled with a gray albedo of 0.1 overimposed to the water ice signature, and located at 45 AU from both the Sun and JWST. Colored regions denote the 3σ (99.7%) confidence interval for each estimate. For example, a 600-km object with 20% water ice, when observed with NIRSpec, would have a determined water-ice fraction of (20±3)%, to 3σ. Simulations based on inputs from A. Guilbert-Lepoutre.

At $R$=1000 various hydrocarbon species can be easily detected, and those are stable against escape on most KBOs and are expected to be present. Such data would also be useful for characterizing volatile species such as $CH_4$, CO, $CO_2$ and $N_2$, but those ices probably only exist on the largest KBOs which are bright enough to be observed at high-SNR in $R$=2700 mode. Sedna is, perhaps, the exception because it is so distant and faint.

Figure 23 illustrates this more clearly, and shows the kinds of absorption features present in the spectra of the largest KBOs. Methane and water ices dominate the spectra, but other species are present albeit with weak and/or narrow absorptions. Silicates are commonly used to model the red visible slopes and spectral shapes near 1 μm and have been detected at 10 μm for at least one centaur object. High-order hydrocarbons similar to those seen in Titan's atmospheric hazes are also frequent components in such models, and are expected photolysis products when $CH_4$ and $N_2$ are present. Of the objects shown in Figure 23, only Pluto has a measured spectrum in the L and M bands. This illustrates just how difficult it is to observe KBOs in the near-IR from the ground: JWST has the sensitivity to revolutionize our understanding of KBO compositions by providing data in this poorly-explored but important spectral region.

Another example of how NIRSpec could be used to distinguish between different ice compositions is shown in Figure 25. The "noisy" spectrum has been generated using a simple simulation of an observation of an Orcus-like object with NIRSpec in its IFU mode and at $R$=1000. In three exposures of 970 s, the signal-to-noise ratio in the spectrum is already high enough to have clear diagnostics on the presence of hydrocarbon, nitrogen and other molecules, as can be seen in the 1.45 – 1.5 µm and 1.65 – 1.8 µm regions of the figure, as well as the crystalline state of the water ice itself (as revealed in the band shape near 1.57 µm). It is important to note that NIRSpec observations would be extremely useful even if they were confined to the narrow wavelength band of Figure 25, but in fact will allow access to longer wavelengths where spectral features from many molecules of interest are much stronger (as can be seen in Figure 23, for example). Thus NIRSpec will be extremely sensitive to details of the surface composition of small bodies, and will allow us to study those compositions at new wavelengths.

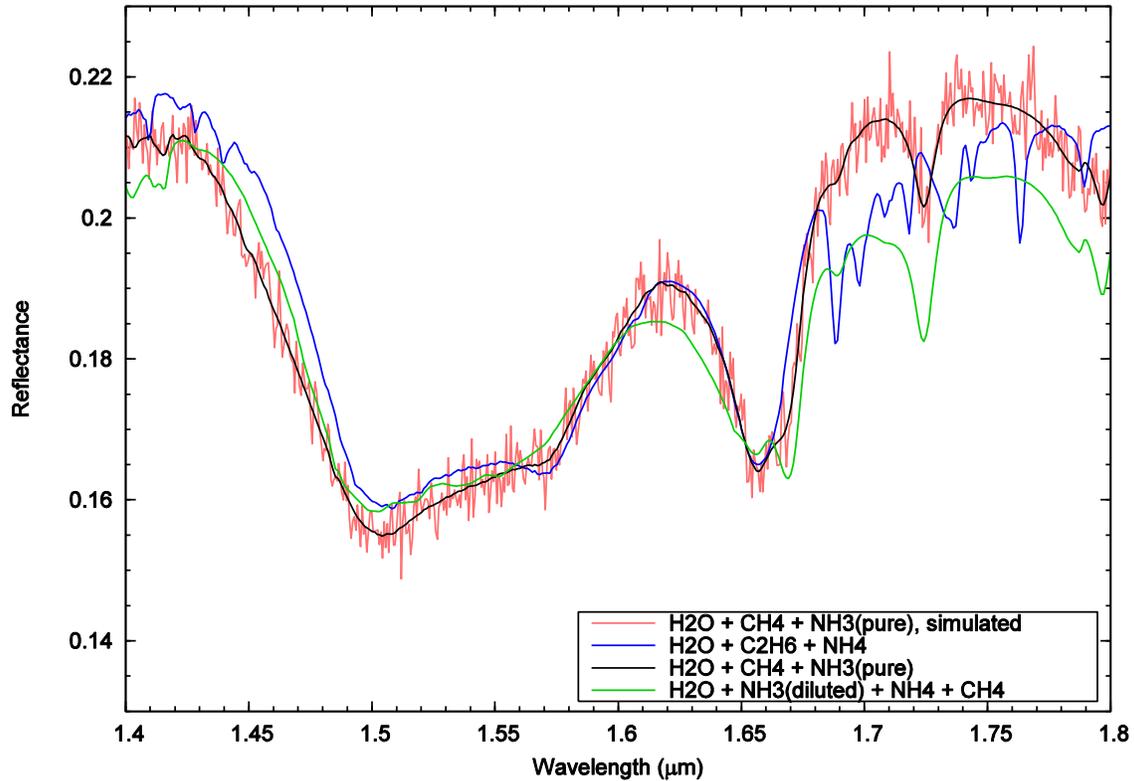

Figure 25. Results of a (simple) simulation of the measurement of the reflectance of an Orcus-like object with NIRSpec in its IFU mode and at medium spectral resolution (F100LP/G140M configuration). Shown here is the 1.4-1.8 µm wavelength range with the reflectance spectra of several different ice mixtures superimposed, including the mixture used as an input for the simulation. Data courtesy of A. Guilbert-Lepoutre.

MIRI will also allow characterization of KBOs, with reflected sunlight dominating the spectra shortward of about 12 µm and thermal emission dominating longward of that, as shown in Figure 26. Photometry at 7.7, 10, 11.3 and 12.8 µm could be diagnostic of silicate or PAH features. Photometry at 18, 21, and 25 µm is diagnostic of the temperature of the KBOs.

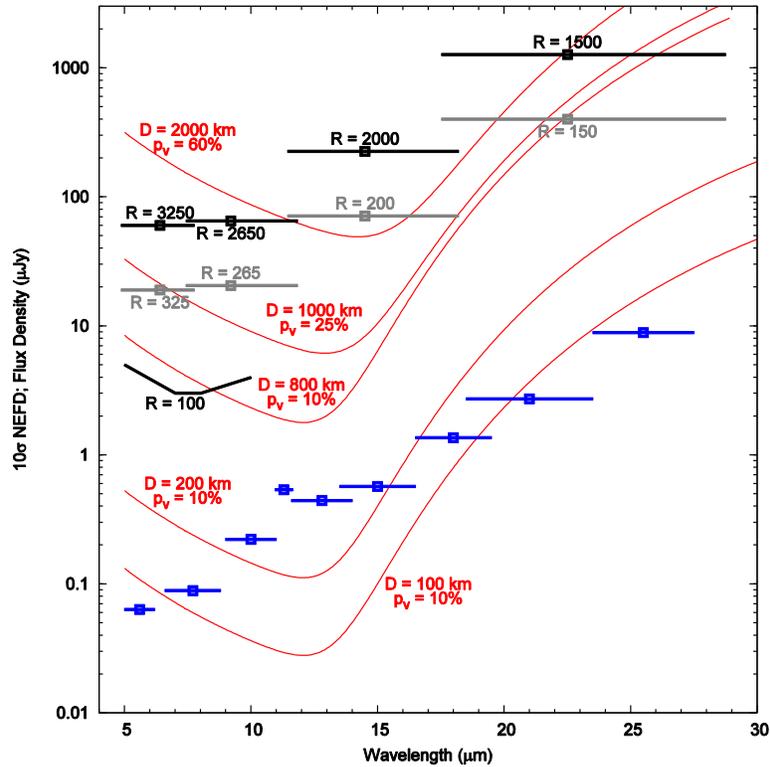

Figure 26. The sensitivity of MIRI spectroscopy (black/gray) and imaging (blue) is compared to model 1000-second spectra for various KBOs (red). Objects are assumed to be located at a distance of 40 AU from both the Sun and JWST, but with different diameters and visual albedos. The spectral sensitivities are plotted for spectra at full resolution (values 1500-3250, plotted in black) and after binning to a resolution ten times lower (plotted in gray). Even with JWST, mid-IR spectra will only be possible for the largest objects. Note that reflected sunlight dominates out to about 12 µm (longer for large objects), while thermal emission is more important longward of that.

Asteroids

Like the comets and KBOs, asteroids are remnants from accretion processes that occurred in the early phases of the formation of the Solar System. Unlike those bodies, asteroids formed inside the "snow-line" of the nebula, interior to which water was present as a gas rather than as ice. Because of that, asteroids are dominated by refractory compounds (silicates, metals, organics), although there is evidence for hydrated minerals (especially in the outer parts of the asteroid belt), and Ceres may contain a significant fraction of water ice in its interior. NIRSpec medium-resolution spectra in the 0.9-5 µm region could be used to search for organics, hydrated minerals, and water ice on a significant sample (~100 objects) of small ($D < 20$ km) asteroids in the outer Main Belt (3.5-4 AU). Features from these materials will occur in the 1.5-5 µm region, as shown in Figure 27; spectra in the 0.9-1.5 µm region will constrain the silicate composition of each body so that a more accurate and complete picture can be drawn of the compositional diversity in general, and between families of asteroids. Current dynamical models for the evolution of the Solar System indicate that some asteroids (including Ceres) may have originated in the Kuiper belt; spectral studies may help in identifying objects with spectra similar to those of KBOs, and distinct from the larger population of asteroids.

JWST provides capabilities for characterizing asteroids that significantly exceed those offered by other facilities in many ways. Wavelength coverage in the critical 2.5–5 µm region, where organics, water and hydrated minerals have strong diagnostic absorptions, combined with the roughly 50 times imaging sensitivity improvement over observatories such as Spitzer, and in many filters, will allow detection of such molecules at lower abundances and on significantly smaller targets than has previously been possible. JWST also provides significant spectroscopic capabilities in this region (which Spitzer did not have and ground-based facilities cannot come close to matching in terms of sensitivity), with spectral resolving powers ranging from ~100 to 2700. The sensitivity of JWST will allow detailed characterization of organic and hydrous molecules, related minor species, and enable searches for previously unidentified molecules on the surfaces of asteroids. JWST will also be 10-50 times more sensitive to thermal emission from asteroids than Spitzer was, and vastly more sensitive than any ground-based facilities. JWST also offers spectroscopy between 5 and 10 microns, which Spitzer did not, and extending to wavelengths of 28.8 µm. JWST measurements in this thermal regime can provide new insights into the thermo-physical characteristics (albedo, roughness, density, thermal inertia) of their surfaces, and of compositional indicators (e.g. silicate and PAH bands in the 10 µm region) that have been impossible using previous or other current facilities. JWST's higher sensitivity will allow the characterization of much smaller and/or more distant asteroids than has been possible. Detailed characterization of significant samples of asteroids will be possible, and could enable, for example, a statistically robust determination of the characteristics of individual asteroid families that could address the heterogeneity (or homogeneity) of the parent bodies of those families.

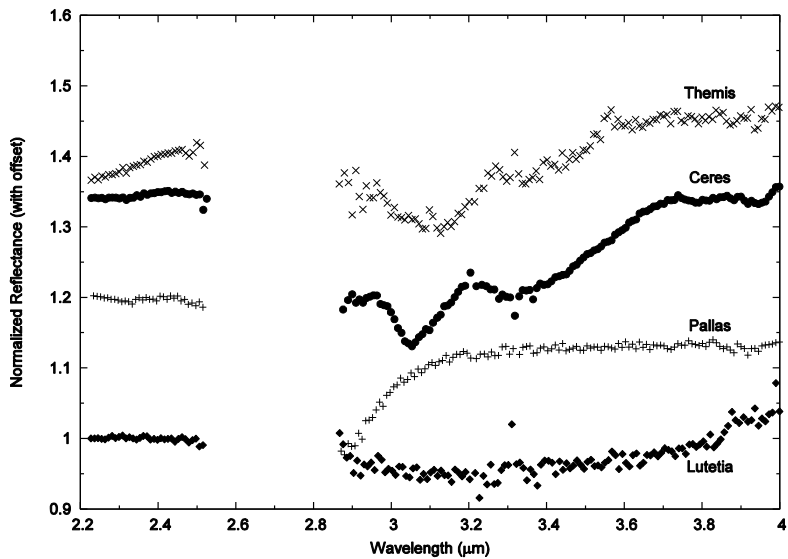

Figure 27. The reflectance spectra of asteroids in the 2-5 µm region display considerable diversity, as illustrated in these ground-based observations (courtesy A. Rivkin). The absorption edge at ~3 µm in the Pallas spectrum is thought to be caused by the 2.7 µm absorption feature of clay minerals. The ~3.1 and ~3.4 µm absorptions in the Themis spectrum arise from water ice and organics, respectively. Ceres displays similar features, but the differences from Themis point to the presence of magnesium hydroxide (~3.05 µm) and coarbonates (~3.3 and ~3.8-3.9 µm). The extremely broad 2.9 - 3.9 µm absorption on Lutetia may be due to goethite. JWST will be able to characterize such absorptions at much higher SNR, for much smaller and fainter targets, and without the spectral-coverage gaps inherent to ground-based spectra.

At MIRI wavelengths asteroid spectra are dominated by thermal emission. Photometric or spectroscopic measurements of that emission can be fit using thermal models to determine asteroid diameters, surface temperatures, albedos, and temperature distributions. Figure 32 illustrates this by giving example thermal-emission spectra for small Main Belt asteroids with a range of thermo-physical properties. The objects were selected to have similar sizes (as determined by IRAS), and are in two groups: high-albedo objects near 2.5 AU (S-type), and low-albedo objects near 3.5 AU (C-type). Asteroids also exhibit spectral features in the mid-IR, produced by silicates such as olivine and pyroxene, and these can be readily detected in MIRI LRS spectra. Figure 29 illustrates the kinds of spectral features we might expect, using emissivity spectra from three Jovian trojan asteroids as examples. Existing mid-infrared data for large Main Belt asteroids show similar features; JWST will allow such observations on much smaller and more diverse objects.

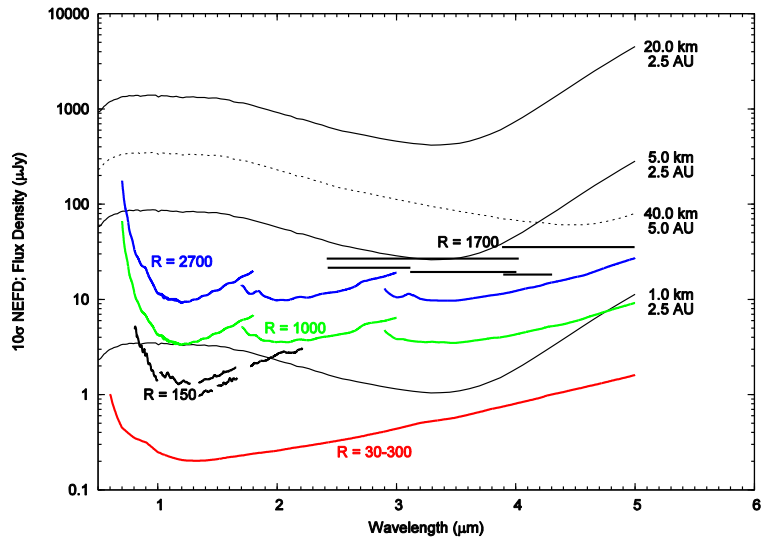

Figure 28. JWST spectral sensitivity (noise-equivalent flux density, or NEFD) curves compared to models of reflected plus emitted spectra for asteroids. The sensitivity curves are for a 1000-second exposure and a signal-to-noise ratio of 10. Thick solid lines are for NIRSpec at spectral resolving powers $30 < R < 300$ (prism), 1000, and 2700. Thick dashed lines are for the NIRISS $R$=150 slitless grism (paired with several different bandpass filters) and for the slitless NIRCam $R$=1700 slitless grism, also paired with available bandpass filters. The thin curves are the model spectral distributions for Main-Belt asteroids (solid lines) and a Jupiter trojan (dashed line), at appropriate distances from the Sun and JWST (assumed equal). The assumed diameter for each object is given in the legend. Objects are assumed to be spectrally neutral, with a visual geometric albedo of 5%.

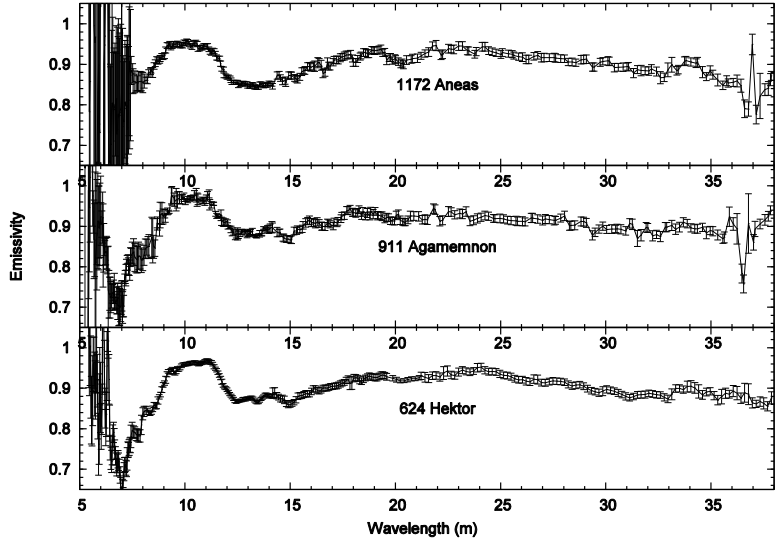

Figure 29. Emissivity spectra of three Jovian trojan asteroids obtained with the Spitzer IRS instrument (Emery et al. 2006). Emission features near 10, 18, 23, and 34 μm are attributed to fine-grained silicates in a transparent matrix or in a fluffy structure. These objects had flux densities in the range 0.1 – 4 Jy over the 10 – 38 μm range (compare with MIRI sensitivity in Figure 30).

MIRI LRS spectra of Main Belt asteroids could also provide sensitive determination of the temperature distribution on their surfaces, as well as compositional information through the silicate emission features broadly clustered around 10 μm wavelength. Each target will be observed twice in order to better constrain the thermal inertia of surface materials. For objects with large-amplitude (>0.25 mag) rotational lightcurves, the two observations might be timed to coincide with lightcurve minimum and maximum. For objects with smaller lightcurve amplitudes (0.1-0.2 mag), one observation will be timed to view the dawn-side emission, and a second later will view the dusk-side emission.

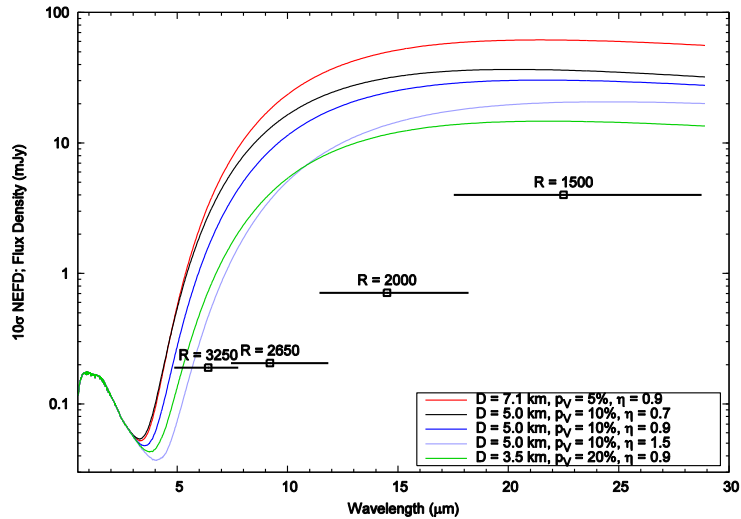

Figure 30. The sensitivity of MIRI spectroscopy for a 1000-second exposure at the highest available resolution is compared to model spectra for small Main Belt asteroids (assumed to be spectrally gray and located at 2.5 AU from both the Sun and JWST). The models are constructed in such a way that all five objects have the same brightness in the visible, but have different diameters, geometric albedos ($p_V$), and surface temperature distributions (as characterized by the beaming parameter $\eta$). MIRI spectroscopy and/or photometry of asteroids will allow characterization of their compositions and physical parameters at very high SNR.

Observations to Complement Other Missions

JWST will also contribute observations that will complement data obtained from numerous spacecraft missions. While JWST cannot compete with the spatial resolution attainable with spacecraft visitations, its large platform has facilitated instrumentation with superior spectral resolution. Furthermore, the longer temporal baseline granted by JWST observations will greatly assist investigations of temporal variations of spacecraft targets, due to changes in season, heliocentric distance, solar activity, or internal processes.

Cassini

NIRSpec and MIRI provide improvements over Cassini's instrument package, offering greater spectral resolution ($R=2300$ and $R=3000$) than VIMS ($R\sim200$) and CIRS ($R \leq 2800$). This will allow a more precise identification and characterization of spectral features in the near and mid infrared. Additionally, MIRI will cover the 5-7 µm gap between the spectral coverage of VIMS and CIRS. Further details regarding JWST observations complementing the Cassini dataset are discussed in earlier sections.

Another important benefit of JWST observations will be more complete coverage of the Saturnian seasonal cycle. Cassini arrived at Saturn in mid-2004, just after southern solstice, and the mission will be terminated in 2017, near northern solstice. JWST will become operational soon afterwards, enabling views of the Saturnian system throughout most of the remaining half of the seasonal cycle: from northern solstice to southern solstice, as shown in Figure 31. A clear benefit will be investigation of the effects of changing sub-solar latitude upon Saturn's atmosphere, rings, and moons, especially to quantify any lag or asymmetry in the seasonal changes. Observing campaigns will also benefit from changing views of the Saturnian system during JWST's lifetime, as different latitudes of Saturn and its moons become sunlit (and therefore visible).

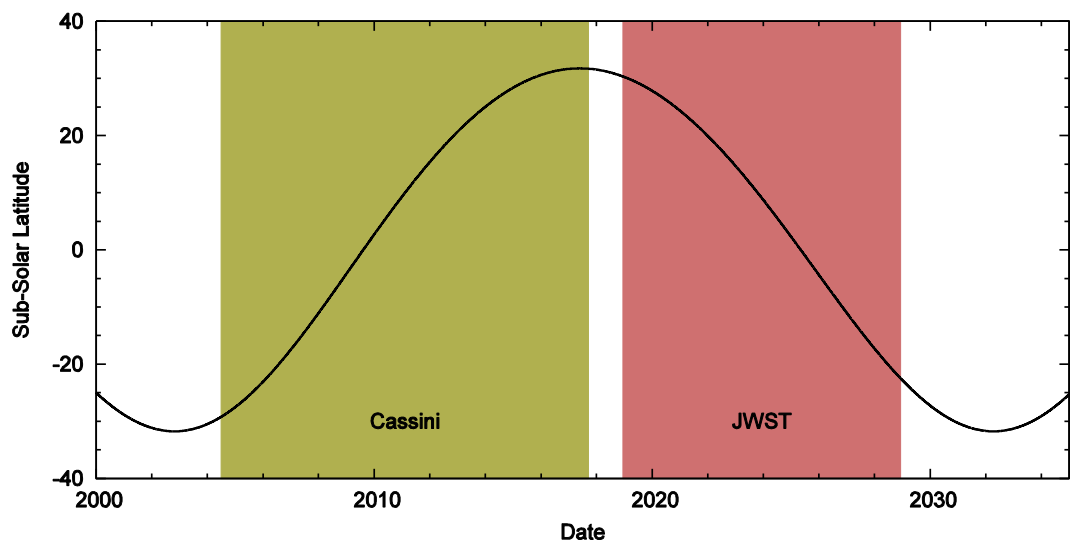

Figure 31. Saturn's sub-solar latitude during the period of JWST operations (red) compared to Cassini operations (gold).

Rosetta

ESA's Rosetta mission was designed to rendezvous with Jupiter-family comet 67P/Churyumov-Gerasimenko. It arrived in mid-2014 when the comet was near 4 AU from the Sun and remained through the comet's August 2015 perihelion, and is expected to end operations in December 2015. As shown in Figure 32, JWST will be able to conduct follow-up observations during the comet's next two orbits.

Few limitations exist in JWST's ability to observe the comet. In previous apparitions, the comet's peak brightness has been V~12[3], so saturation will not be an issue. The comet's 2-km nucleus (Kelley et al. 2009) is also large enough to be observed near aphelion with NIRSpec (using moderate resolution) and MRS. The only issue will be the comet's speed: for a period near the comet's 2021 perihelion, it will be moving too fast for JWST to track.

---

[3] http://sci.esa.int/science-e/www/object/index.cfm?fobjectid=14615

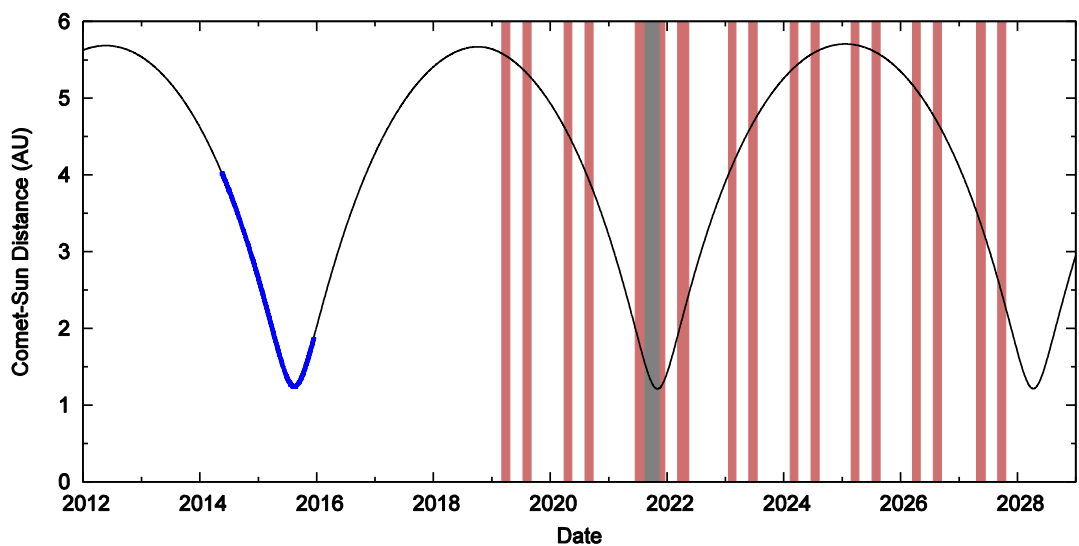

Figure 32. Observations of comet 67P/Churyumov-Gerasimenko compared to its heliocentric distance. Rosetta's 2014-2015 rendezvous with the comet is shown as a thick blue line. JWST observing windows are shown as solid red bars. The gray region near the comet's 2021 perihelion denotes where the comet is at a viewing angle observable with JWST, but moving faster than JWST's maximum tracking rate of 0.03″ per second. The comet's 2028 perihelion occurs during a broad span where it is at an angle not observable with JWST.

New Horizons

During the July 2015 flyby of Pluto by New Horizons, instruments onboard the spacecraft studied the Pluto system primarily in the ultraviolet, visible, near infrared (out to 2.5 μm), and radio. As such, JWST observations of these worlds will entail almost entirely different parts of the spectrum, contributing to the larger picture of Pluto and its moons. JWST data will also expand upon the temporal context of New Horizons observations, providing insight regarding any evolution (on sub-seasonal timescales) that was ongoing at the time of the flyby.

While JWST will be able to distinguish Pluto from its moons in imagery, these objects are each too small to be resolved. However, the detailed mapping conducted by New Horizons (in addition to the slow rotation rates of Pluto and Charon) will enable JWST users to clearly identify which surface features are in view at a given time. Such information may help identify the source of any unusual spectral signatures, or allow studies of a particular feature through comparisons of when it is visible and rotated out of view.

Pluto itself is too small in angular size to be resolved in JWST imagery, but it is bright enough for near-infrared spectroscopy with NIRSpec at maximum resolution of $R$=2700. In spectral regions that overlap with the New Horizons spectrometer LEISA, this is a tremendous improvement over LEISA's maximum spectral resolution of $R$=600. Charon, at ~25% the angular area of Pluto, should be similarly observable in most of the near-infrared, though lower-resolution spectroscopy (e.g., $R$=1000) may be necessary in regions with particularly strong absorption features.

Spectroscopy of Pluto will be possible in parts of the mid-infrared. At full resolution, MRS will be able to obtain spectra shortward of 6 μm and longward of 22 μm; binning to a resolution 10 times lower would add additional narrow regions to the spectrum, as shown in Figure 26. Even with such binning, spectra of Charon would only be possible shortward of ~6 μm and longward of ~23 μm. However, both objects are well above the sensitivity limits for all of MIRI's imaging filters, so photometry will be an option where spectroscopy is not.

Regarding Pluto's four tiny moons—Nix, Hydra, Kerberos, and Styx—photometry will be possible using some NIRCam filters. The larger pair of moons, Nix and Hydra, will be observable in wide filters shortward of 4 μm, and in four medium filters between 1.3 and 2.2 μm (Figure 22). The four tiny moons will be impossible to observe in the mid-infrared, with even the largest satellites' anticipated brightnesses falling short of the sensitivity limits of the MIRI imager (Figure 26).

Other Missions

There are several other missions where seasonal changes are less important, but JWST will still be able to augment the observational dataset. None of the missions listed below have mid-infrared capabilities, and JWST will have much greater spectral resolution in the near infrared. Near- and mid-infrared observations of these mission targets with JWST will greatly assist in precise characterization of global surface properties of the spacecraft targets, as well as enable investigations of their thermal properties.

- NASA's Dawn mission, having orbited Vesta in 2011-2012, arrived at Ceres in March 2015, and continues to conduct observations at the time of this writing. Other upcoming asteroid visitations include OSIRIS-REx, currently scheduled to investigate the asteroid 101955 Bennu over 2018-2021; and Hayabusa2, which will explore the asteroid 162173 Ryugu over 2018-2019. The targets of both of these missions will be observable with JWST.

- The mission timespan for JWST (2018-2028) almost completely spans the interim between the Juno and JUICE missions, operated by NASA and ESA, respectively. Juno will be at Jupiter from July 2016 to October 2017. JUICE will reach Jupiter in January 2030 at the earliest, and will observe Jupiter and the three outer Galilean moons for almost three years before entering orbit around Ganymede.

**Online Resources**

Below are links to JWST pages online. These pages are routinely updated with current information.

JWST design and instrument overview:

http://www.stsci.edu/jwst/overview/design/

Resources for Solar System observations with JWST:

http://www.stsci.edu/jwst/science/solar-system

PDF "pocket guides" for each instrument:

MIRI: http://www.stsci.edu/jwst/instruments/miri/docarchive/miri-pocket-guide.pdf

NIRCam: http://www.stsci.edu/jwst/instruments/nircam/docarchive/NIRCam-pocket-guide.pdf

NIRSpec: http://www.stsci.edu/jwst/instruments/nirspec/docarchive/NIRSpec-pocket-guide.pdf

NIRISS: http://www.stsci.edu/jwst/instruments/niriss/docarchive/NIRISS-pocket-guide.pdf

TABLES

Table 1. Movement Rates of Various Solar System Targets.

| Object | Minimum Rate (mas/sec) | Maximum Rate (mas/sec) |
|---|---|---|
| Mars | 0.485 | 28.27 |
| Ceres | 0.152 | 11.81 |
| Jupiter | 0.019 | 4.48 |
| Saturn | 0.016 | 1.74 |
| Uranus | 0.012 | 1.09 |
| Neptune | 0.020 | 0.74 |
| Pluto | 0.004 | 0.65 |
| Haumea | 0.372 | 0.62 |
| Eris | 0.058 | 0.30 |

Table 2. Estimated Surface Brightness of Outer Planets and Pluto.

| Planet | Heliocentric Distance (AU) | Radius (km) | Albedo | 2 μm (Jy/☐″) | 4-5μm | 10-12 μm | Diameter (″) | Area (☐″) |
|---|---|---|---|---|---|---|---|---|
| Mars | 1.67 | 3397 | 0.2 | 100 | 34 | 2800 | 7.0 | 39 |
| Jupiter | 5.2 | 71492 | 0.52 | 22 | 100 | 35 | 39 | 1172 |
| Saturn | 9.5 | 60268 | 0.47 | 6.0 | 2 | 10 | 18 | 243 |
| Uranus | 19.2 | 25559 | 0.51 | 1.6 | 0.47 | 0.3 | 3.7 | 11 |
| Neptune | 30.1 | 24766 | 0.41 | 0.5 | 0.16 | 100 | 2.3 | 4 |
| Pluto | 33.0 | 1150 | 0.55 | 0.0055 | 0.0016 | 0.0003 | 0.10 | 0.008 |

Note: These values are the sum of reflected solar flux and thermal emission. Angular sizes assume each target is at a solar elongation of 90°, and therefore within the JWST field of regard. Mars is assumed to be near aphelion, when it is at its dimmest. Jupiter's surface brightness is for a Jovian "hot spot," which is brighter than the disk average. Pluto's brightness values are for full disk (in Jy), rather than surface brightness.

Table 3. Observing Methods for Outer Planets with JWST Instruments.

| Planets | NIRCam | NIRSpec | MIRI Imager | MRS |
|---|---|---|---|---|
| Mars | With the $160^2$ sub-array (wherein the full disk will fit as in Fig. 1), observable in F210M, F300M, F480M, and all narrow filters. Remaining medium filters and some wide filters become possible with the $64^2$ sub-array. | Observable with the 0.2″ × 3.3″ slit at long wavelengths (>2.9 μm) using a standard spectral window; shortward requires non-standard spectral windows (0.70-0.75 μm, 2.65-3.0 μm). IFU useable only in certain non-standard windows near 0.7, 2.8, and 4.3 μm. | Above saturation limits, even with smallest sub-arrays. | Above saturation limits, even with smallest sub-arrays. |
| Jupiter | Sub-array imaging not required for narrow filters and a few medium filters. Using the $640^2$ sub-array (as in Fig. 1) will make most medium filters useable; smaller sub-arrays will make all filters possible. | IFU useable at most wavelengths, but may saturate near 0.9 and 1.1 μm. Jupiter will be fully below the saturation limits for the slit. Note that both apertures are much smaller than Jupiter, so specific features should be targeted, or mosaicking should be used. | Darker regions with thicker ammonia cloud observable in F580W and possibly F770W, with SUB64 sub-array. (Note that SUB64 is very small compared to the Jovian disk.) Hot spots are above saturation limits. | May be possible in some spectral regions shortward of 10 μm. |
| Saturn | All narrow, and most medium filters longward of 1.75 μm, will not require sub-arrays. The $400^2$ sub-array, which fits the Saturnian disk as in Fig. 1, will make all medium filters and some wide filters useable; the rest will require the smaller $160^2$ sub-array, which has only a 5″ field of view in the SW channel. | Both slit and IFU are useable at all wavelengths, though the field of view of either is smaller than the size of the planet. Mosaicking will be required to produce full-disk spectral maps. | SUB64 enables mid-IR imaging shortward of 16 μm, though full-planet views will require mosaicking. | Likely possible at wavelengths shortward of ~17 μm. |
| Uranus and Neptune | Neptune is observable in all filters without sub-arrays; Uranus is similar but may require sub-arrays for F070W and F080W. Both planets are well above NIRCam sensitivity limits even in the darkest spectral regions. | No saturation issues. The IFU is ideal, as it is comparable in size to both planets. | Without subarrays, Neptune is observable in four filters, and Uranus in all but the two longest-wavelength filters. If the SUB64 subarray is used, all filters are useable for both planets. SUB64 is larger in size than these planets' disks. | Both planets well below the saturation limits. Uranus is rather dim over most of 5.0-8.5 μm; binning or the LRS ($R$~100) may be preferred. MIRI IFUs are similar in size to the planets' disks. |

Note: All spectroscopic observations discussed assume maximum-resolution modes (NIRCam: $R$=2700, MRS: $R$~1550-3250), which also provide the highest saturation limits. SUB64 denotes the smallest sub-array for the MIRI imager, with a field of view of 7.0″ × 7.0″.

Table 4. Estimated Minimum Sizes of Detectable Satellites.

| Planet | Optimal Filter | Satellite Diameter (km) |
|--------|----------------|-------------------------|
| Jupiter | F1500W | 0.3 |
| Saturn | F1500W | 1.5 |
| Uranus | F2100W | 9 |
| Neptune | F2550W | 30 |

Note: Satellite sizes are estimated for SNR=10 after 10,000 seconds of integration with MIRI. The satellites are assumed to follow the Standard Thermal Model, with low albedos such that $(1 - a)^{1/4} \approx 1$. Background signal was estimated using the Exposure Time Calculator, assuming maximum zodiacal light brightness. Due to the gradually changing nature of a satellite's predicted brightness relative to the background, filters adjacent to the optimal filter will provide similar results.